%% file: hierarchy_in_press.tex
\documentclass[12pt]{article}

\usepackage{isomath}
\usepackage{amsmath}
\usepackage{amsbsy}
\usepackage{amssymb}
\usepackage{amscd}
\usepackage{amsfonts}
\usepackage{graphicx,color}
\usepackage{verbatim}
\usepackage{euscript}
\usepackage[parfill]{parskip}
\usepackage{alltt}
\usepackage{stmaryrd}
\usepackage[utf8]{inputenc}
\usepackage[english]{babel}
\usepackage{fancyhdr}
\usepackage{xy}
\usepackage[margin=1in]{geometry}
\usepackage{caption}
\usepackage{url}
\usepackage[titletoc,title]{appendix}
\usepackage{float}
\restylefloat{table}
\usepackage{setspace}
\usepackage{xcolor}
\usepackage{enumitem}
\usepackage{breqn}
\usepackage[round]{natbib}
\usepackage{setspace}

\title{How useful are formal hierarchies? A case study on averaging dislocation dynamics to define meso-macro plasticity}

\author{Sabyasachi Chatterjee\thanks{Dept.\ of Civil \& Environmental Engineering, Carnegie Mellon University, Pittsburgh, PA 15213. sabyasac@andrew.cmu.edu.} 
\and Amit Acharya\thanks{Dept.\ of Civil \& Environmental Engineering, and Center for Nonlinear Analysis, Carnegie Mellon University, Pittsburgh, PA 15213. acharyaamit@cmu.edu.}}

\input{latexConfigureFile}

\graphicspath{{./imagedir/}}

\date{}
\allowdisplaybreaks
\begin{document}
\maketitle

\begin{abstract}
\noindent A formal hierarchy of exact evolution equations are derived for physically relevant space-time averages of state functions of microscopic dislocation dynamics. While such hierarchies are undoubtedly of some value, a primary goal here is to expose the intractable complexity of such systems of nonlinear partial differential equations that, furthermore, remain `non-closed,' and therefore subject to phenomenological assumptions to be useful. It is instead suggested that such hierarchies be terminated at the earliest stage possible and effort be expended to derive closure relations for the `non-closed' terms that arise from the formal averaging by taking into account the full-stress-coupled microscopic dislocation dynamics (as done in \cite{ddfdm_coupling}), a matter on which these formal hierarchies, whether of kinetic theory type or as pursued here, are silent.
\end{abstract}

\colorlet{change}{blue}

\newcommand{\ePm}{ e(-P^m) }
\newcommand{\eQn}{ e(-Q^n) } 
\newcommand{\PmMinusOne} { P^{m-1} }
\newcommand{\QnMinusOne} { Q^{n-1} }

\newcommand{\Pdot} { {\left(\frac{ ({\bfalpha} {\bfn}^l)\cdot ( -curl({\bfalpha} \times {\bfV})^T) {\bfn}^l}  {c_1 |{\bfalpha} {\bfn}^l|} \right)}}

\newcommand{\alphaTildeDot} {\frac{curl({\bfalpha}\times {\bfV})}{|\bfalpha|} - \left( \frac{ {\bfalpha}:curl({\bfalpha}\times{\bfV}) }{ |\bfalpha|^3} \right) ~ \bfalpha}

\newcommand{\alphaTildeTranpDot} {\frac{\{curl({\bfalpha}\times{\bfV})\}^T}{|\bfalpha|} - \left( \frac{ {\bfalpha}:\{curl({\bfalpha}\times{\bfV})\} }{ |\bfalpha|^3} \right) ~ {\bfalpha}^T}

\newcommand{\alphaTildeBarDotOne} {\Big\{ {-curl({\color{blue}\overline{\bfalpha}} \times {\color{blue}\overline{\bfV}} + {\bfL}^p )}  ~\overline{ \Big({1 \over \bfalpha}\Big)} + \overline{ {\Sigma}^{-curl(\bfalpha \times \bfV)} {\Sigma}^{1 \over {|\bfalpha|} } } \Big\}   } 

\newcommand{\alphaTildeBarDotTwo} {\Big[ \Big\{ \Big( -{\color{blue}\overline{\bfalpha}} : {curl({\color{blue}\overline{\bfalpha}} \times {\color{blue}\overline{\bfV}} + {\bfL}^p )} + \overline{{\Sigma}^{\bfalpha} : {\Sigma}^{-curl({\bfalpha}\times {\bfV})}} \Big) \overline{ \Big({1 \over {|\bfalpha|^3}}\Big)}}
\newcommand{\alphaTildeBarDotThree} {  \overline{{\Sigma}^{ -{\bfalpha}:{curl({\bfalpha} \times {\bfV})}} {\Sigma}^{1 \over {|\bfalpha|^3}} } \Big\} {\color{blue}\overline{\bfalpha} }
+ \overline{ {\Sigma}^{\frac{ -{\bfalpha}:curl({\bfalpha} \times {\bfV}) }{ |\bfalpha|^3} } {\Sigma}^{\bfalpha} } \Big]}

\newcommand{\alphaTildeTranspBarDotOne} {\Big\{ ({-curl({\color{blue}\overline{\bfalpha}} \times {\color{blue}\overline{\bfV}} + {\bfL}^p )})^T ~ \overline{ \Big({1 \over \bfalpha}\Big)} + \overline{ {\Sigma}^{-(curl(\bfalpha \times \bfV))^T} {\Sigma}^{1 \over {|\bfalpha|} } } \Big\}   }

\newcommand{\alphaTildeTranspBarDotTwo}{ \Big[ \Big\{ \Big( -{\color{blue}\overline{\bfalpha}} : { (curl({\color{blue}\overline{\bfalpha}} \times {\color{blue}\overline{\bfV}} + {\bfL}^p )}) + \overline{{\Sigma}^{\bfalpha^T} : {\Sigma}^{-(curl({\bfalpha} \times {\bfV})^T)}} \Big) \overline{\Big({1 \over {|\bfalpha|^3}}\Big)}  }

\newcommand{\alphaTildeTranspBarDotThree} {  \overline{{\Sigma}^{ -{\bfalpha}:{(curl({\bfalpha} \times {\bfV}))}} {\Sigma}^{1 \over {|\bfalpha|^3}} } \Big\} \overline{\bfalpha^T} + \overline{ {\Sigma}^{\frac{ -{\bfalpha}:(curl({\bfalpha} \times {\bfV})) }{ |\bfalpha|^3} } {\Sigma}^{\bfalpha^T} } \Big]}

\newcommand{\QdotOne} {sgn(p)  ~ \tilde{{\bfb}}^l \cdot \Big[ \Big\{ -\frac{curl({\bfalpha}\times {\bfV})}{|\bfalpha|} + \big( \frac{ {\bfalpha}:curl({\bfalpha}\times{\bfV}) }{ |\bfalpha|^3} \big)  ~ \bfalpha \Big\} \tilde{\bfalpha}^T }

\newcommand{\QdotTwo} {\tilde{\bfalpha} \Big\{ -\frac{\{curl({\bfalpha}\times{\bfV})\}^T}{|\bfalpha|}  + \big( \frac{ {\bfalpha}:\{curl({\bfalpha}\times{\bfV})\} }{ |\bfalpha|^3} \big)  ~ {\bfalpha}^T \Big\} \Big] \tilde{{\bfb}}^l }

 \newcommand{\PbarDotFirst} {-{1 \over {c_1}} \Big( {\color{blue}\overline{ {\bfalpha}}} {\bfn}^l  \cdot \{{curl( {\color{blue}\overline{\bfalpha}} \times {\color{blue}\overline{\bfV}} + {\bfL}^p) {\bfn}^l} \}}

 \newcommand{\PbarDotSecond} { \overline{ {\Sigma}^{{\bfalpha} {\bfn}^l} \cdot {\Sigma}^{{ -curl({\bfalpha} \times {\bfV}) {\bfn}^l  }} } \Big)  \overline{\left( 1 \over {|{\bfalpha} {\bfn}^l}|\right)}  
+ \overline{ {\Sigma}^{-({\bfalpha} {\bfn}^l) \cdot { (curl({\bfalpha} \times {\bfV}) {\bfn}^l)}}   ~ {{\Sigma}^{1 \over {|{\bfalpha} {\bfn}^l}|}  } } }

\newcommand{\QbarDotOne} { \overline{sgn(|p|-1)} ~\overline{ sgn(p) } ~ \tilde{{\bfb}}^l \cdot {\color{brown}\Bigg\{ } {\color{cyan}\Bigg( }{\alphaTildeBarDotOne}}
\newcommand{\QbarDotTwo} {  {\alphaTildeBarDotTwo} }
\newcommand{\QbarDotThree} { {\alphaTildeBarDotThree} {\color{cyan}\Bigg)} ~ \overline{\tilde{\bfalpha}^T} }
\newcommand{\QbarDotFour} { {\overline{\tilde{\bfalpha}}} ~ {\color{magenta}\Bigg(} {\alphaTildeTranspBarDotOne}}
\newcommand{\QbarDotFive} { {\alphaTildeTranspBarDotTwo} }
 \newcommand{\QbarDotSix} { ~ {\alphaTildeTranspBarDotThree} {\color{magenta}\Bigg)} {\color{brown}\Bigg\} }\tilde{{\bfb}}^l}
 \newcommand{\QbarDotSeven}  { \overline{sgn(|p|-1)}~ \overline{ \, {\color{purple}\Bigg( } \Sigma^{sgn(p)}  \Sigma^{\tilde{{\bfb}}^l \cdot {\color{teal}\Big[ }{ \big(-\frac{curl({\bfalpha}\times {\bfV})}{|\bfalpha|} - \left( \frac{ {\bfalpha}:curl({\bfalpha}\times{\bfV}) }{ |\bfalpha|^3} \right) ~ \bfalpha \big) 
 \tilde{\bfalpha}^T } } } } 
\newcommand{\QbarDotEight}  { \qquad ~ \qquad ~\qquad ~ \overline{ {}^{ + \tilde{\bfalpha} \Big( -\frac{\{curl({\bfalpha}\times{\bfV})\}^T}{|\bfalpha|} - \left( \frac{ {\bfalpha}:\{curl({\bfalpha}\times{\bfV})\} }{ |\bfalpha|^3} \right) ~ {\bfalpha}^T  \Big) {\color{teal} \Big] } \tilde{{\bfb}}^l } {\color{purple}\Bigg) } \, } } 
\newcommand{\QbarDotNine} {  \overline{ {\Sigma}^{sgn(|p|-1)} {\color{red} \Bigg( }  \Sigma^{sgn(p)~ \tilde{{\bfb}}^l \cdot {\color{brown}\Big[} { \Big(-\frac{curl({\bfalpha}\times {\bfV})}{|\bfalpha|} - \left( \frac{ {\bfalpha}:curl({\bfalpha}\times{\bfV}) }{ |\bfalpha|^3} \right) ~ \bfalpha \Big) \tilde{\bfalpha}^T } } }  }
\newcommand{\QbarDotTen } { \qquad ~ \qquad ~ \qquad ~ \overline{ {}^{+ \tilde{\bfalpha} \Big( -\frac{\{curl({\bfalpha}\times{\bfV})\}^T}{|\bfalpha|} - \left( \frac{ {\bfalpha}:\{curl({\bfalpha}\times{\bfV})\} }{ |\bfalpha|^3} \right) ~ {\bfalpha}^T  \Big) {\color{brown}\Big]} \tilde{{\bfb}}^l}  {\color{red} \Bigg) } \, } } 
 \newcommand{\chidotOne} {m ~{e{{\left(\frac{|{\bfalpha} {\bfn}^l |} {c_1}\right)}^m}} ~ {e{{\left(  \frac{|{\tilde{\bfb}	}^l.\tilde{\bfalpha} \tilde{\bfalpha}^T . \tilde{\bfb}^l|} {c_2}\right) }^n}} ~{{\left(\frac{|{\bfalpha} {\bfn}^l |} {c_1}\right)}^{m-1}} ~ \Big( \frac{ ({\bfalpha} {\bfn}^l) \cdot ( \{-curl({\bfalpha} \times {\bfV}) \} {\bfn}^l) } {c_1 ~ |{\bfalpha} {\bfn}^l |} \Big) } 
\newcommand{\chidotTwo} {n ~ {e{(P^m)}} ~ {e{(Q^n)}} ~ {e{(p)}} ~ {p}^{n-1}} 
\newcommand{\chidotThree} { sgn(p) ~ \tilde{\bfb}^l \cdot \Big\{ \Big(\frac{-curl({\bfalpha} \times {\bfV})}{|\bfalpha|} - ( \frac{ {\bfalpha}:-curl({\bfalpha} \times {\bfV}) }{ |\bfalpha|^3} ) ~ \bfalpha \Big) \tilde{{\bfalpha}^T} }
\newcommand{\chidotFour} { + \tilde{\bfalpha} \Big(\frac{-(curl({\bfalpha}\times{\bfV}))^T}{|\bfalpha|} - ( \frac{ {\bfalpha}:-(curl({\bfalpha}\times{\bfV}))^T }{ |\bfalpha|^3} ) ~ {\bfalpha}\Big) \Big\} \tilde{\bfb}^l}

\newcommand\sbullet[1][.5]{\mathbin{\vcenter{\hbox{\scalebox{#1}{$\bullet$}}}}}

\section{Introduction}\label{sec:intro}
This paper is concerned with the formal derivation of governing field equations of increasingly detailed space-time averaged behavior of microscopic dislocation dynamics, and assessing the value of such systems. The microscopic dislocation dynamics is posed as a system of pde, capable of representing the dynamics of a collection of possibly tangled smooth curves representing dislocation core cylinders, each core cylinder movable by a combination of glide and climb due to the action of a vectorial velocity field. The velocity field is determined, following well-accepted notions, purely from the dislocation density field (with the possibly tangled web of core cylinders viewed simply as appropriate smooth localizations in space of the dislocation density field), and the (nonlinear crystal elastic) stress field in the body; even when linear elasticity is used, the point-wise Burgers vector direction and line direction (information built into the dislocation density field) are adequate to describe the motion of edge segments, and the motion of screws are restricted to within a geometrically defined set of planes. In any case a resolution into slip system dislocation densities is not essential (cf. \cite{zhu_chapman_ach}).  This pde system is adequate for representing the plasticity of the constituent material when atomic length scales are resolved - we refer to this system as Field Dislocation Mechanics (FDM). We are interested in obtaining the implications of this model when the resolved length (and time) scales are much coarser, i.e. we are interested in obtaining some information on the nature of the governing evolution equations for increasingly detailed descriptions of averaged behavior of this microscopic system, appropriate for coarser-length and time scales. We emphasize that the derived averaged equations represent exact, but non-closed, statements of evolution of the defined average variables, without any compromise on the inherent kinematic constraints of the microscopic system (e.g. the connectedness of the dislocation lines represented by the solenoidal property of the microscopic dislocation density field).

The above line of inquiry was initiated in \cite{acharya_roy_2006, acharya_chapman}; as will be shown in this paper, the exact equations of evolution become exceedingly complex and cumbersome and it was suggested in \cite{acharya_roy_2006} that closure assumptions be made at a relatively lower level to maintain tractability (while allowing for the inclusion of all that is known in the physics-based phenomenological modeling of plastic deformation and strength, e.g. \cite{kocks_argon_ashby}) and refining the description as required for greater fidelity. We will refer to this approach as the MFDM (Mesoscale Field Dislocation Mechanics) approach to plasticity.

In the Continuum Dislocation Dynamics (CDD) framework of Hochrainer and collaborators \cite{hochrainer2007three, hochrainer2016thermodynamically, zaiser_monavari},  models are developed based on a kinetic theory-like framework, starting from the assumption that a fundamental statement for the evolution of a number density function on the space of dislocation segment positions and orientations is available at the microscopic level. This \textit{microscopic} governing equation is non-closed even if one knows completely the rules of physical evolution of individual dislocations segments of connected lines; one would need to study the behavior of an ensemble of dislocation dynamics evolutions to define, and then also only in principle, the evolution of such a number density function (cf. \cite[Sec 3.1, 5]{hochrainer2007three}) - this detail is built into the state-space velocity function introduced in \cite{hochrainer2007three}, which cannot be simply defined by a well-accepted statement like the Peach-K\"{o}hler force for a segment of a real-space description of a dislocation line. Furthermore, Equations (7) and (11) of \cite{hochrainer2007three}, the fundamental statement of evolution governing the number density function (a `collective' quantity), are postulated without fundamental justification. This is in contrast to FDM where the \textit{fundamental justification for the statement of microscopic dynamics} is the integral statement of conservation of Burgers vector, a physically observed fact (which does not imply a conservation of the `number' of dislocations, whether loops or otherwise, as stated in \cite[Sec. 4]{hochrainer2007three}, and as demonstrated in exercises related to annihilation and nucleation \cite{garg_ach_maloney}); then, the equations of MFDM follow strictly from FDM on averaging, without any further assumptions. Returning to CDD, on making various assumptions for tractability, the theory produces (non-closed) statements of evolution for the averaged dislocation density (akin to the mesoscale Nye tensor field), the total dislocation density (similar to an appropriate sum of the averaged Nye tensor density) and, these densities being defined as physical scalars, and an associated curvature density field. Closure assumptions are made to cut off infinite hierarchies, which is standard for averaging based on nonlinear `microscopic equations', and further closure assumptions for constitutive statements are made based on standard thermodynamic arguments \cite{hochrainer2016thermodynamically}.

The model in \cite{xia_elazab} belongs to the same mathematical class as MFDM  but with more complicated constitutive structure related to multiple-slip behavior \cite{das2016microstructure}. It assumes geometrically linear kinematics for the total deformation coupled to a system of stress-dependent, nonlinear transport equations for vector-valued slip-system dislocation densities. These slip system density transport equations involve complicated, phenomenological constitutive assumptions related to cross-slip, and the authors of \cite{xia_elazab} promote the point of view that dislocation patterning is related to the modeling of slip system dislocation densities and cross-slip. An attempt to understand emergence of microstructure in 1-d setting was made in \cite{das2016microstructure}, and in 3-d and finite deformation setting was made in \cite{arora2020dislocation}. They concluded that in all likelihood, such complexity is not essential for the emergence of dislocation microstructure in this family of models. 

Despite \textit{postulating} the microscopic dynamics on which their entire approach is based, making phenomenological closure assumptions of their own, and making ad-hoc choices of cutting off the infinite hierarchy of their equations, the CDD authors have criticized the MFDM approach as inadequate for describing the plasticity of metals \cite{hochrainer2007three, hochrainer2016thermodynamically, zaiser_monavari}. The authors of \cite{xia_elazab} have criticized MFDM for using a phenomenological approach and for the lack of resolution into slip systems in order to model cross slip. One goal of this paper is to show that the criticisms leveled against the MFDM approach in \cite{hochrainer_materials, xia_elazab, zaiser_monavari} are unfounded, at least in comparison to the standards of these works.

The MFDM approach starts from a well-accepted fundamental microscopic dynamical statement (unlike CDD), and produces an exact hierarchy of equations for any desired level of detail in the coarse description as an implication of this fundamental microscopic dynamics. Extending the work in \cite{acharya_chapman}, this paper explicitly shows that while the MFDM approach can easily accommodate descriptors like slip system dislocation densities and define precise hierarchies of evolution equations for them, such an enterprise comes at a significant cost in complexity and tractability of the resulting model, and shows the exact nature of phenomenology and gross approximation that would necessarily be inherent in any proposed formalism for coarse-grained dislocation dynamics (e.g., \cite{hochrainer_materials,xia_elazab}) that does not consider head-on the question of averaging the stress-coupled interaction-related dynamics of dislocations.

This paper is organized as follows. In Section \ref{sec:hierarchy}, we apply the averaging procedure utilized in \cite{babic1997} to generate an infinite hierarchy of nonlinear coarse equations corresponding to a fine dynamics, which (essentially) cannot be solved. In Section \ref{sec:conn_with_mfdm}, we apply the averaging procedure to derive the coarse evolution of averaged total dislocation density and demonstrate, using a simple example, that the averaged dislocation density of an expanding circular loop increases. In Section \ref{subsec:refined_desc}, we present a refined description of the variables with respect to individual slip systems, reflective of a crystal plasticity description and present the coarse evolution of these variables, namely the averaged dislocation density tensor and the averaged total dislocation density. In order to do so, we define a characteristic function, which indicates whether any given position has a dislocation of a particular slip system. We show how the coarse evolution of these variables are very cumbersome and hence, why it is reasonable to close the infinite hierarchy of non-closed system of equations at a low level. It is also important to generate lower level closure assumptions that account for the \textit{stress coupled} dynamics of dislocations. Such work has been demonstrated in \cite{ddfdm_coupling}.

\section{Hierarchy of averaged equations for nonlinear microscopic equations: the basic idea}\label{sec:hierarchy}
In this section, we will utilize an averaging procedure used in the literature for multiphase flows (see \cite{babic1997}). For a microscopic field $f$ given as a function of space and time, the mesoscopic space-time-averaged field $\bar{f}$ \cite{acharya_roy_2006,acharya_chapman} is given as  
\begin{align}\label{eq:avg}
\bar{f}({\bfx},t)={1 \over {\int_{I(t)} \int_{\Omega({\bfx})} w({\bfx}-{\bfx}', t-t') d{\bfx}'dt'}} {\int_{\Im} \int_{B} w({\bfx}-{\bfx}', t-t') f({\bfx}',t')d{\bfx}'dt'},
\end{align}
where $B$ is the body and $I$ a sufficiently large interval of time. In the above, $\Omega({\bfx})$ is a bounded region within the body around the point ${\bfx}$ with linear dimension of the order of the spatial resolution of the macroscopic model we seek, and $I(t)$ is a bounded interval in $I$ containing $t$. The weighting function $w$ is non-dimensional, assumed to be smooth in the variables ${\bfx}$, ${\bfx}'$, $t$, $t'$ and, for fixed ${\bfx}$ and $t$, have support (i.e. to be non-zero) only in $\Omega({\bfx}) \times I(t)$ when viewed as a function of $({\bfx}',t')$. 

The one-dimensional analogue of \eqref{eq:avg} is
\begin{align}\label{eq:heir_pde}
{\partial_{t'} f} = F(f,{\partial_{x'} f}),
\end{align} 
where $x'$ is the spatial coordinate and $t'$ is time and $f$ is a function of $x'$ and $t'$. We call the system given by equation \eqref{eq:heir_pde} as the fine scale system. 

We aim to understand the macroscopic evolution of the fine dynamics \eqref{eq:heir_pde} in terms of averaged (coarse) variables. To do so, the averaging operator \eqref{eq:avg} is applied to both sides of \eqref{eq:heir_pde}, which results in the following:
\begin{align}\label{eq:heir_d1}
\frac{\partial}{\partial{t}} \overline{f}(x,t) = \overline{F(f,{\partial_{x'} f})} (x,t).
\end{align}

We denote $A_0:=\overline{f}$ and $A_{01}:={\partial_t A_0}=\overline{F}$. The fluctuation of function $f$ is defined as: 
\begin{align}\label{eq:fluc}
{\Sigma}^f (x',t',x,t):=f(x',t')-\overline{f}(x,t). 
\end{align}

The average of the product $(p)$ of two variables $f$ and $g$ is given by
\begin{align}\label{eq:product_avg_genl}
\overline{f (p) g} =& \overline{ \{\bar{f} + (f - \bar{f}) \} (p) \{  \bar{g} + (g - \bar{g}) \}}  \nonumber\\
= &\overline{ \bar{f} (p) \bar{g} + \bar{f} (p) (g-\bar{g}) + (f-\bar{f}) (p) \bar{g}  + (f-\bar{f}) (p) (g- \bar{g})} \nonumber\\
= & \overline{f} (p) \overline{g} + \bar{f} (p) \overline{\Sigma^g} +  \bar{g} (p) \overline{\Sigma^f} + \overline{ {\Sigma}^f (p) {\Sigma}^g} \nonumber\\
=& \overline{f} (p) \overline{g} + \overline{ {\Sigma}^f (p) {\Sigma}^g},
\end{align}
so that 
\begin{align*}
\overline{f(p)g}-\bar{f}(p)\bar{g}=\overline{f(p)g-\bar{f}(p)\bar{g}}=\overline{ {\Sigma}^f (p) {\Sigma}^g}.
\end{align*}
Here, $f$ and $g$ can be scalar, vector or tensor valued. Some examples of the product (p) are scalar multiplication, vector inner product, tensor inner product, cross product of a tensor with a vector etc. For example, if $f$ and $g$ are scalar and $(p)$ is the scalar multiplication operator, 
\begin{align}\label{eq:product_avg_scalar}
\overline{f g} = \overline{f} \, \overline{g} + \overline{ {\Sigma}^f  {\Sigma}^g}.
\end{align}

Using \eqref{eq:product_avg_genl}, we obtain the average of product of three variables as
\begin{align}\label{eq:product_avg_3vars_genl}
\overline{f(p)g(p)h}=&\overline{f}(p) \overline{g(p)h} + \overline{\Sigma^f (p) \Sigma^{g(p)h}}=\overline{f}(p) \{ \overline{g}(p)\overline{h} + \overline{\Sigma^g (p) \Sigma^h} \} + \overline{\Sigma^f (p) \Sigma^{g(p)h}}\nonumber\\
=& \overline{f}(p)\overline{g}(p)\overline{h} + \overline{f} (p) \overline{\Sigma^g (p) \Sigma^h} + \overline{\Sigma^f (p) \Sigma^{g(p)h}}.
\end{align}

Similarly, for $f$, $g$ and $h$ scalars and $(p)$ the scalar multiplication operator, 
\begin{align}\label{eq:product_avg_3vars_scalar}
\overline{f g h} = \overline{f} \, \overline{g} \, \overline{h} +  \overline{f} \, \overline{\Sigma^g \, \Sigma^h} + \overline{\Sigma^f \, \Sigma^{g \, h}}.
\end{align}

Equations \eqref{eq:product_avg_scalar}, \eqref{eq:product_avg_3vars_genl} and \eqref{eq:product_avg_3vars_scalar} show that the \emph{averages of products of two or more variables are not the products of their averages}.


It can be shown using \eqref{eq:heir_pde}, \eqref{eq:heir_d1} and \eqref{eq:fluc} that 
\begin{align}\label{eq:heir_d2}
{\partial_{tt} A_0}={\partial_t A_{01}}=\overline{{\partial_2 F}{\partial_1 F}} {\partial_x A_0} + \overline{{\partial_2 F}{\partial_2 F}} {\partial_{xx} A_0} + A_{011} + \overline{{\Sigma}^{({\partial_2 F}{\partial_1 F})} {\Sigma}^{({\partial_{x'} f})}} + \overline{{\Sigma}^{({\partial_2 F}{\partial_2 F})} {\Sigma}^{({\partial_{x'x'} f})}},
\end{align}
where $A_{011}:=\overline{{\partial_1} FF}$. 



Thus, in the coarse evolution of $\bar{F}$, new terms (e.g. $\overline{{\partial_2 F}{\partial_1 F}}$) emerge and therefore, it is necessary to augment \eqref{eq:heir_d2} with coarse evolution equations of the new terms (that appear on the rhs of \eqref{eq:heir_d2}), namely the following:  
\begin{align}\label{eq:G_i}
{\partial_t \left(\overline{{\partial_2 F}{\partial_1 F}}\right)}&=G_1 \nonumber\\
{\partial_t \left(\overline{{\partial_2 F}{\partial_2 F}}\right)}&=G_2 \nonumber\\
{\partial_t \left(\overline{{\Sigma}^{({\partial_2 F}{\partial_1 F})} {\Sigma}^{({\partial_{x'} f})}}\right)}&=G_3 \nonumber\\
{\partial_t \left(\overline{{\Sigma}^{({\partial_2 F}{\partial_2 F})} {\Sigma}^{({\partial_{x'x'} f})}}\right)}&=G_4,
\end{align}
where $G_i$ ($i=1$ to $4$) are functionals of the state. In general, these functionals cannot be expressed as functionals of the independent fields of the averaged model, such relations being referred to as `closure' relations. 

The functionals $G_i$ can be generated by applying the averaging operator \eqref{eq:avg} to the fine scale equations and decomposing the average of product of functions in the fine scale into product of their averages and their associated fluctuation terms. For example, the coarse evolution of $\overline{{\partial_2 F}{\partial_1 F}}$ (which is given by $G_1$ above) is 
\begin{align*}
{\partial_t \left(\overline{{\partial_2 F}{\partial_1 F}}\right)}&= \overline{\partial_t' \left({\partial_2 F}{\partial_1 F}\right)}= \overline{\partial_t' \left({\partial_2 F}\right){\partial_1 F}+ {\partial_2 F} \partial_t' \left( \partial_1 F \right)} \\
&= \overline{H({\partial_1 F})+ ({\partial_2 F})M} = \overline{H} \overline{\partial_1 F} + \overline{M} \overline{\partial_2 F} + \overline{ \Sigma^H \Sigma^{\partial_1 F} } +  \overline{ \Sigma^M \Sigma^{\partial_2 F} },
\end{align*}
where $H=\partial_t' \left({\partial_2 F}\right)$ and $M=\partial_t' \left( \partial_1 F \right)$. Thus, new terms appear in the coarse evolution of the term $\overline{{\partial_2 F}{\partial_1 F}}$, and similarly for the other terms in \eqref{eq:G_i}. In this manner, we can generate an infinite hierarchy of nonlinear coarse equations corresponding to the fine dynamics \eqref{eq:heir_pde}. Solution of such an infinite system is not possible. 

It is therefore necessary to close the equations at a desired level, which means to use physics based assumptions for the necessary terms instead of solving their exact evolution equation (for example, in equation \eqref{eq:G_i} above, we  might use closure assumptions for the functionals $G_i$ on the rhs). Morever, even if the system was finite but large and could be solved (in principle), approximating solutions to nonlinear systems of pde is by no means a trivial task, so that it is definitely better to shift the focus from generating large formal hierarchies of nonlinear pde to generating controlled, with respect to accuracy,  closure assumptions to maintain tractability.

\section{Models of MFDM with varying coarse descriptors} \label{sec:conn_with_mfdm}
%
The model of FDM \cite{acharya_2001,acharya_2003,acharya_2004} represents the dynamics of a collection of dislocation lines at the atomic length scale. The field equations of FDM are as follows:
\begin{align}\label{eq:FDM}
&\dot{\bfalpha}=-curl({ \bfalpha}\times {\bfV}) \nonumber\\
&curl {\bfchi}={\bfalpha}\nonumber\\
&div{\bfchi}=0 \nonumber \\
&div(grad\dot{\bfz})=div({\bfalpha}\times{\bfV} + {\bfL}^p) \nonumber \\
&div({\bfC}:\{grad{\bfu}-{\bfz} + {\bfchi}\})=0.
\end{align}
The tensor $\bfalpha$ is the dislocation density tensor, ${\bfV}$ is the dislocation velocity vector, ${\bfC}$ is the fourth-order, possibly anisotropic, tensor of linear elastic moduli, ${\bfu}$ is the total displacement vector, $\bfchi$ is the incompatible part of the elastic distortion tensor, and ${\bfu}-{\bfz}$ is a vector field whose gradient is the compatible part of the elastic distortion tensor. Upon application of the averaging operator \eqref{eq:avg} defined in Section \ref{sec:hierarchy} to both sides of \eqref{eq:FDM}, we have the following system of averaged equations
\begin{align}\label{eq:MFDM}
&\dot{\overline{\bfalpha}}=-curl(\overline{\bfalpha}\times\overline{\bfV}+{\bfL}^{p})\nonumber\\
&curl\overline{\bfchi}=\overline{\bfalpha}\nonumber\\
&div \overline{\bfchi}=0 \nonumber\\
&div(grad\dot{\overline{\bfz}})=div(\overline{\bfalpha}\times \overline{\bfV} + {\bfL}^p) \nonumber\\
&div(\overline{\bfC}:\{grad(\overline{\bfu}-\overline{\bfz}) + \overline{\bfchi} \})=0
\end{align}
\cite{acharya_roy_2006}. The system \eqref{eq:MFDM} is called \emph{Mesoscale Field Dislocation Mechanics}. Here, ${\bfL}^p$ is defined as 
\begin{align}\label{eq:Lp}
{\bfL}^p:=\overline{{\bfalpha}\times{\bfV}} - \overline{\bfalpha}\times \overline{\bfV},
\end{align}
and it represents the strain rate produced by `statistically stored dislocations'. It follows from \eqref{eq:product_avg_genl} that ${\bfL}^p$ is the average of the cross product of the fluctuation of $\bfalpha$ and $\bfV$ ( which means $\bfL^p=\overline{\Sigma^{\bfalpha} \times \Sigma^{\bfV}}$). Consider a uniformly expanding square loop. Since ${\bfalpha}={\bfb}\otimes{\hat{\bfl}}$, where $\bfb$ is the Burgers vector density per unit area and $\hat{\bfl}$ is the line direction at each point of the loop and $\bfb$ remains uniform along the loop, and both $\hat{\bfl}$ and $\bfV$ change sign going from one side of the square loop to the opposite side,  both $\overline{\bfalpha}={\bf0}$ and $\overline{\bfV}={\bf0}$. However, ${\bfalpha}\times{\bfV}$ is identical for opposite sides of the loop and does not cancel out and hence, ${\bfL}^p \neq {\bf0}$. 

\subsection{Isotropic MFDM}\label{sec:iso_mfdm}
We consider as descriptors of the system the averaged total dislocation density $\rho$ and the plastic distortion rate $\bfL^p$, which are commonly used in the literature (also see \cite{acharya_chapman}). 

\subsubsection {Evolution equation for \textbf{averaged total dislocation density}, $\overline{\rho^l}$.} The total dislocation density is defined as 
\begin{align}\label{eq:rho}
\rho:={\bfalpha}:{\bfalpha}. 
\end{align}

Suppose we have many dislocation segments in a box of volume $V$. We see that $\frac{\int_{V} \rho dv}{V}=\frac{\sum_i \bfalpha_i : \bfalpha_i ~ l_i A_i}{V}$, where $\bfalpha_i$, $l_i$ and $A_i$ (which is assumed to be $|\bfb_i|^2$ up to a constant) are the dislocation density tensor, line length and cross section area of segment $i$ respectively. We also have that $\bfalpha_i =\frac{|\bfb_i| \bfm_i \otimes \bft_i}{A_i}$, where $\bfb_i$, $\bfm_i$ and $\bft_i$ are the Burgers vector, Burgers vector direction and the line direction of segment $i$. Therefore, $\frac{\sum_i \bfalpha_i : \bfalpha_i ~ l_i A_i}{V}={1 \over V} \sum_i \frac{|\bfb_i|^2}{A_i^2}l_i A_i={1 \over V} \sum_i \frac{|\bfb_i|^2}{|\bfb_i|^4}l_i |\bfb_i|^2={1 \over V} \sum_i l_i$, which is the averaged dislocation density in the box. Since $\rho$ is the microscopic total dislocation density, $\frac{\int_{ V} \rho dv}{V}$  is $\rho$ averaged over $V$, which gives the averaged dislocation density of the box.  This acts as a verification that $\rho$ is indeed the total dislocation density.

\begin{figure}[!h]
\centering
  \includegraphics[width=0.35\linewidth]{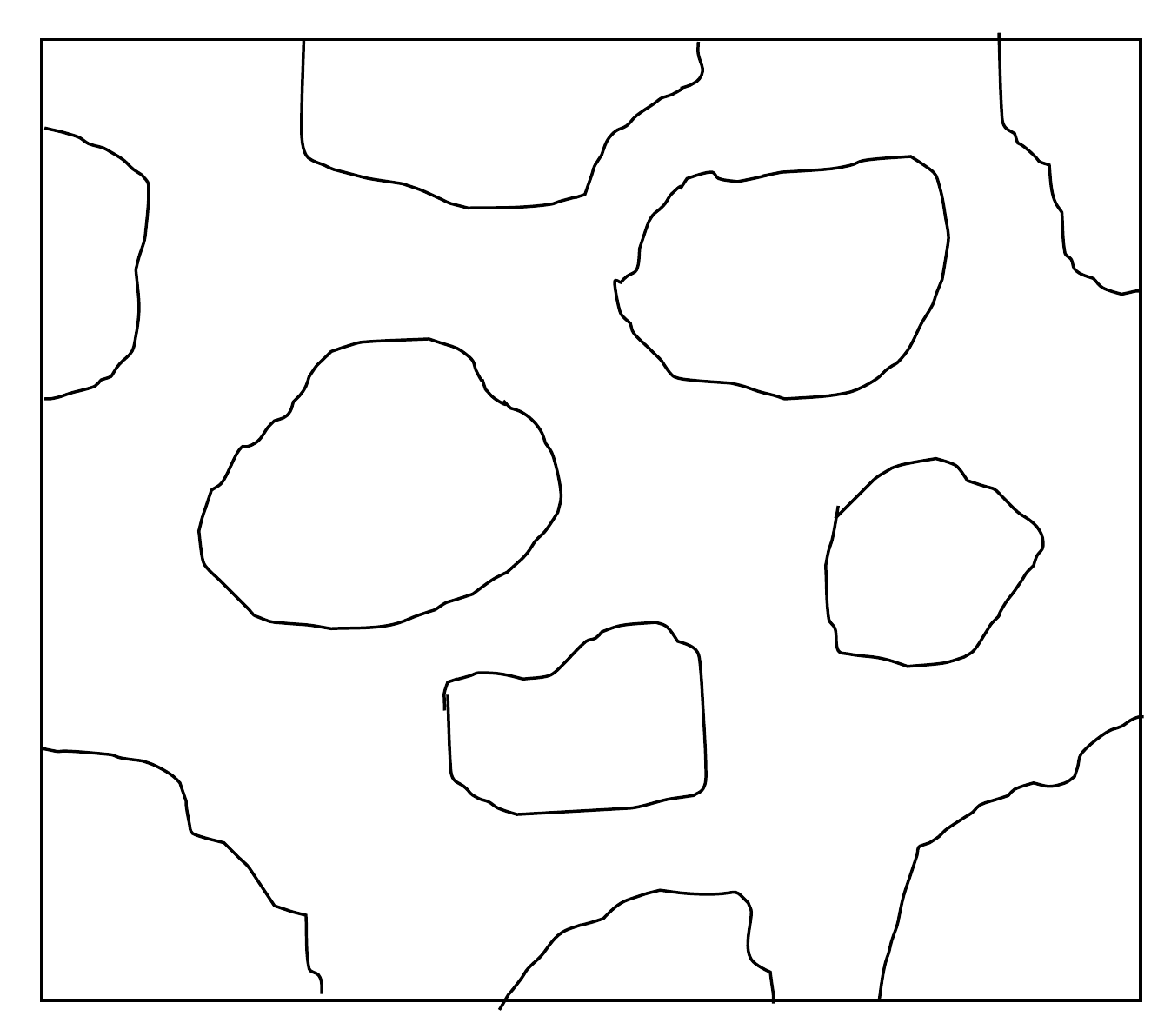}
  \caption{\textit{$Dislocation~ loops~ in~ averaging~ box$} }
  \label{fig:loops}
\end{figure}

The space-time averaged total dislocation density $\overline{\rho}$ is given by 
\begin{align}\label{eq:rho_avg}
\overline{\rho}=\overline{\bfalpha}:\overline{\bfalpha} + \overline{ {\Sigma}^{\bfalpha} : {\Sigma}^{\bfalpha} }.
\end{align}  
This follows from \eqref{eq:product_avg_genl} and shows that the average of the total dislocation density contains average terms as well as averages of fluctuations. We can interpret this using Fig. \ref{fig:loops} in which we see that the averaging box has many loops which are inside the box and there are some loops which are not entirely contained inside the box. Since the Burgers vector is uniform over a loop, the average dislocation density (${\bfb} \otimes \hat{\bfl}$) due to the loops which are contained in the box is $\bf0$ since the average of the line direction $\hat{\bfl}$ over the loop cancels out. The only contribution to the first term on the rhs of \eqref{eq:rho_avg} is from the loops which are not entirely contained in the averaging box. If our averaging box has a very large length scale, then most loops will be contained inside the box and as such, $\overline{\bfalpha} \approx {\bf0}$ and the main contribution to the averaged total dislocation density will come from the average of the fluctuation term given by the second term on the rhs of \eqref{eq:rho_avg}.  The evolution of such fluctuation terms, as discussed in Section \ref{sec:hierarchy}, will be given by other pde, which will themselves be non-closed, as they will contain other fluctuation terms. This will generate an infinite hierarchy of non-closed cumbersome pde, as will be shown next.  

The evolution of $\overline{\rho}$ is given by
\begin{align}\label{eq:rhobardot}
\dot{\overline{\rho}}=& - grad~ \overline{\rho} \cdot \overline{\bfV} - 2 ~\overline{\rho}~ div \overline{\bfV} + 2 ~\overline{\bfalpha} : ( div \overline{\bfalpha} \otimes \overline{\bfV} ) + 2~ \overline{\bfalpha}: \{ \overline{\bfalpha} ~ grad \overline{\bfV} \} - \overline{ {\Sigma}^{grad {\rho}} \cdot {\Sigma}^{\bfV}}  \nonumber \\
& - 2 \overline{{\Sigma}^{\rho} {\Sigma}^{div V}} + 2 ~\overline{\bfalpha} : ( \overline{ \Sigma^{div \bfalpha}  \otimes \Sigma^{\bfV}} ) + 2 \overline{ \Sigma^{\bfalpha} : \Sigma^{ div \bfalpha \otimes \bfV} } +  2~ \overline{\bfalpha} : \overline{ {\Sigma}^{\bfalpha} ~ {\Sigma}^{grad{\bfV}} } \nonumber\\
& + 2~ \overline{ {\Sigma}^{\bfalpha} :  {\Sigma}^{ {\bfalpha} ~ {grad{\bfV}} }  }.   
\end{align}
The derivation of \eqref{eq:rhobardot} is given in Appendix \ref{subsec:rhobardot}. As the averaging length scale becomes large, the RHS of \eqref{eq:rhobardot} is dominated by the averages of the fluctuation terms.

\textbf{Example: Circular dislocation loop} 

\begin{figure}[!h]
\centering
  \includegraphics[width=0.35\linewidth]{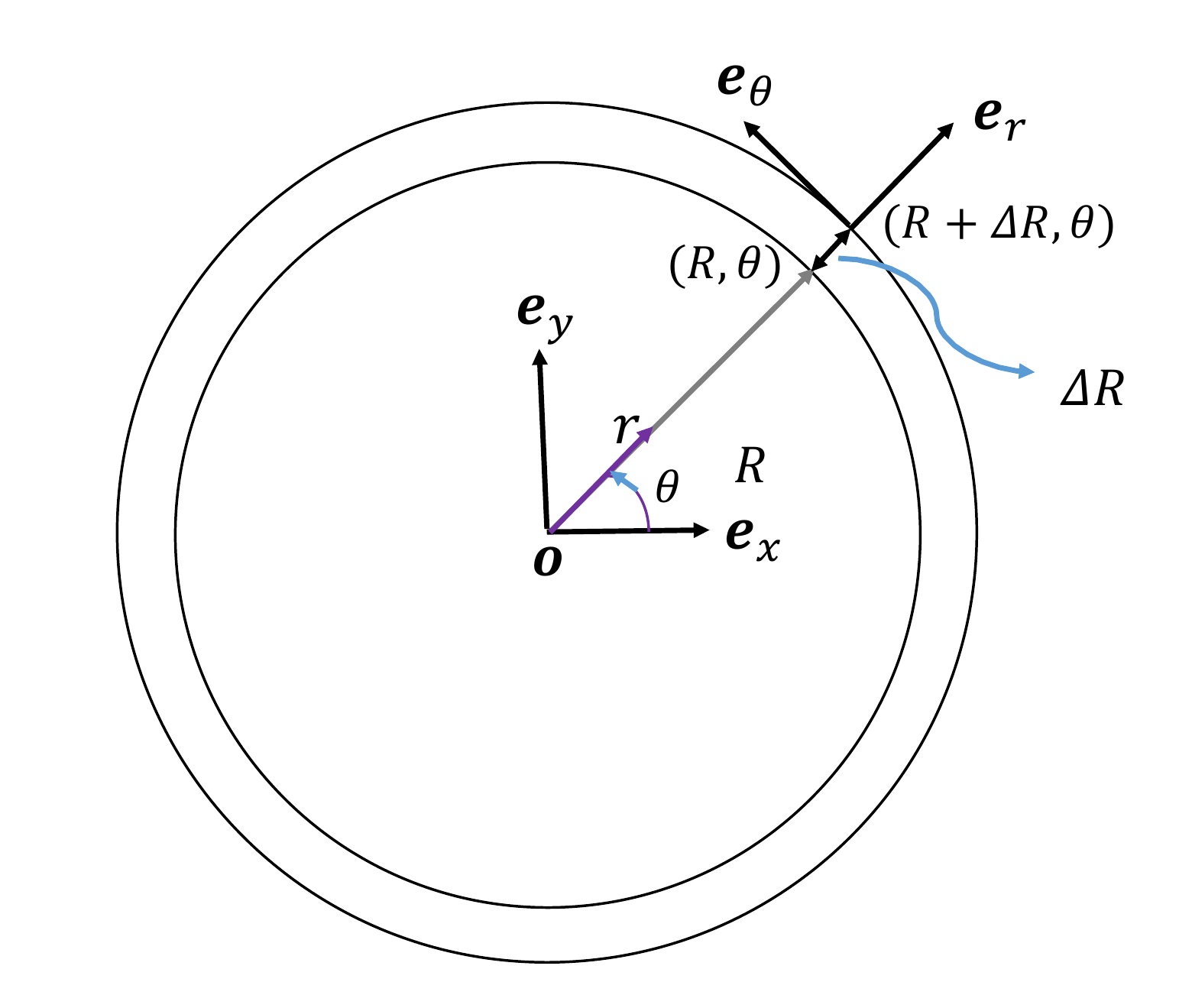}
  \caption{\textit{Top view of a uniformly expanding loop of radius $R$ and width $\Delta R$.} }
  \label{fig:exp_loop}
\end{figure}

The evolution equation for $\rho$ (as derived in Appendix \ref{subsec:rhobardot} and given by \eqref{eq:rhodot}) is 
\begin{align*}
\dot{\rho}&= -grad \rho \cdot {\bfV} - 2 \rho (div V) +2 \bfalpha: (div \bfalpha \otimes \bfV) + 2 {\bfalpha}: \{{\bfalpha} \,grad V\} 
\end{align*}
Application of the averaging operator \eqref{eq:avg} to the above results in the following: 
\begin{align}\label{eq:rhobardot_chap}
\dot{\overline{\rho}}=-\overline{grad \rho \cdot {\bfV} } - \overline{2 \rho (div V)} + \overline{2 \bfalpha: (div \bfalpha \otimes \bfV)} + \overline{2 {\bfalpha}: \{{\bfalpha} \,grad V\}}. 
\end{align} 

We aim to understand the evolution of $\overline{\rho}$ for the case of a circular dislocation loop of inner radius $R$, width $\Delta R$ (see Fig. \ref{fig:exp_loop}) and thickness $t$. The area of cross section of the loop is $A=\Delta R .\, t$, which is assumed to be $b^2$, where $b$ is the magnitude of the Burgers vector $\bfb$. The radial unit vector is $\bfe_r=cos \theta \bfe_x + sin \theta \bfe_y$, while the tangential unit vector is $\bfe_{\theta}=-sin \theta \bfe_x + cos \theta \bfe_y$. Let us assume that its velocity has the same magnitude for all points $(r,\theta,z)$ of the loop (where $z$ is the spatial coordinate along the thickness) and points radially outwards. Hence, the velocity is given by ${\bfV}=v(r)~{\bfe}_r$, where $v(r)=\tilde{v} H(r-R) - \tilde{v} H(r-(R+\Delta R))$. Let the averaging domain be a circular plate of radius $L$ and thickness $H$, where $L \gg R$ and $H \gg t$.

We have 
\begin{equation}\label{eq:vbar}
\begin{split}
\overline{\bfV}=& {1 \over \pi L^2 H} \int_{\theta=0}^{\theta=2\pi} \int_{r=0}^{r=L} \int_{z=-\frac{t}{2}}^{\frac{t}{2}} v(r) {\bfe_r} r dr d\theta dz \\
= & \frac{t}{\pi L^2 H} \int_{\theta=0}^{\theta=2\pi} \int_{r=0}^{r=L} \{\tilde{v} H(r-R) - \tilde{v} H(r-(R+\Delta R))\} \{cos\theta {\bfe_x} + sin\theta {\bfe_y} \} r dr d\theta \\
=&\frac{\tilde{v} t}{\pi L^2 H} \Big[\int_R^{R+\Delta R} r~dr\Big] \Big[ \Big( \int_{\theta=0}^{\theta=2\pi} cos\theta \Big) {\bfe}_x + \Big( \int_{\theta=0}^{\theta=2\pi} sin\theta \Big) {\bfe}_y \Big] \\
= & \frac{\tilde{v} t} {\pi L^2 H} \left[ \frac{r^2}{2} \right]_R^{R+\Delta R} ~ \Big[ \{sin {2 \pi} - sin {\it 0}\} {\bfe}_x +  \{cos {2 \pi} - cos {\it 0} \} {\bfe}_y \Big]\\ 
= & \frac{\tilde{v} t}{\pi L^2 H} ~ \Delta R ~ (2 R + \Delta R) ~{\bf0}=\bf0
\end{split}
\end{equation}

Also, 
\begin{equation}\label{eq:alphabar}
\begin{split}
\overline{\bfalpha}=&{1 \over \pi L^2 H} \int_{\theta=0}^{\theta=2\pi} \int_{r=0}^{r=L} \int_{z=-\frac{t}{2}}^{\frac{t}{2}} \, \frac{\bfb}{A} \otimes \hat{\bfl} r dr d\theta dz = \frac{t}{\pi L^2 H b^2} \int_{\theta=0}^{\theta=2\pi} \int_{r=0}^{r=L} \bfb \otimes \bfe_{\theta} r dr d\theta \\
= & \frac{t} {\pi L^2 H b^2} \int_{\theta=0}^{\theta=2\pi} \int_{r=0}^{r=L} \bfb \otimes  \{-sin\theta {\bfe_x} + cos\theta {\bfe_y} \} r dr d\theta \\
=&\frac{t} {\pi L^2 H b^2} \bfb \otimes \Big[\int_R^{R+\Delta R} r~dr\Big] \Big[ \Big( \int_{\theta=0}^{\theta=2\pi} -sin\theta \Big) {\bfe}_x + \Big( \int_{\theta=0}^{\theta=2\pi} cos\theta \Big) {\bfe}_y \Big] \\
= &\frac{t} {\pi L^2 H b^2} \bfb \otimes \left[ \frac{r^2}{2} \right]_R^{R+\Delta R} ~ \Big[ \{cos {2 \pi} - cos {\it 0}\} {\bfe}_x -  \{sin {2 \pi} - sin {\it 0} \} {\bfe}_y \Big] \\
=&\frac{t} {\pi L^2 H b^2}\bfb \otimes \{ \Delta R . (2 R + \Delta R) .{\bf0} \}=\frac{2 R + \Delta R} {\pi L^2 H}\bfb \otimes {\bf0} =\bf0. 
\end{split}
\end{equation}

Moreover, 
\begin{align*}
grad {\bfV}&= \frac{\partial{V_r}}{\partial r} {\bfe_r} \otimes {\bfe_r} + \frac{V_r}{r} {\bfe_{\theta}} \otimes {\bfe_{\theta}} \\
\implies div \bfV&= grad \bfV:{\bfI} = \frac{\partial{v(r)}}{\partial r} + \frac{v(r)}{r} \\
&=\tilde{v} \delta(r-R) - \tilde{v} ~\delta(r-(R+\Delta R)) + \frac {\tilde{v} ~ [ H(r-R) - H(r-(R+\Delta r))]}{r}
\end{align*}
Hence, 
\begin{equation}\label{eq:divVbar}
\begin{split}
\overline{div({\bfV})}&= {1 \over \pi L^2} \int_{\theta=0}^{\theta=2\pi} \int_{r=0}^{r=L} \int_{z=-\frac{t}{2}}^{\frac{t}{2}}  grad(\bfV) r dr d\theta dz\\
&= \frac{\tilde{v} t}{\pi L^2 H}  \int_{\theta=0}^{\theta=2\pi} \int_{r=0}^{r=L} \Big\{ \tilde{v} ~\delta(r-R) - \tilde{v} . \delta(r-(R+\Delta R)) \\
&\qquad \qquad \qquad + \frac {\tilde{v} ~ [ H(r-R) - H(r-(R+\Delta r))]}{r} \Big\} r dr d\theta \\
&= \frac{\tilde{v} t}{\pi L^2 H}  [R - (R+\Delta R)] . (2\pi) + \frac{\tilde{v} t}{\pi L^2 H}  \Big\{\int_R^{R+\Delta R} dr\Big\} . (2 \pi) \\ 
&= -\frac{\tilde{v} t}{\pi L^2 H} . \Delta R . (2\pi)  + \frac{\tilde{v} t}{\pi L^2 H}  . \Delta R  . (2 \pi) =0
\end{split}
\end{equation}

We also note that $\overline{div{\bfV}}=div{\overline{\bfV}} = div{\textit {\textbf 0}} = {\it0}$. We also have 
\begin{align}\label{eq:gradrho}
\rho=&\bfalpha : \bfalpha = \frac{\bfb}{ A} \otimes \hat{\bfl} : \frac{ \bfb}{ A } \otimes \hat{\bfl} = {1 \over A^2 } (\bfb \cdot \bfb) (\hat{\bfl} \cdot \hat{\bfl}) = \frac{b^2}{ b^2.b^2 } . (1)= {1 \over b^2} \nonumber \\
\implies & grad \rho = \frac{\partial \rho} {\partial r} \bfe_r + {1 \over r} \frac{\partial \rho}{\partial \theta} \hat{\bfe_{\theta}} = 0 + 0 =0.  
\end{align}

We have that 
\begin{align*}
div \bfalpha =& \frac{\partial \bfalpha} {\partial r} \bfe_r + {1 \over r} \frac{\partial \bfalpha}{\partial \theta} \bfe_{\theta} \nonumber\\
=& \frac{\partial (\frac{\bfb}{ A} \otimes \hat{\bfl})}{\partial r} \bfe_r + {1 \over r}  \frac{\partial ( \frac{\bfb}{ A} \otimes \hat{\bfl})}{\partial \theta} \bfe_{\theta} \nonumber \\
=& {1 \over b^2} \bfb \otimes \frac{\partial \hat{\bfl}}{\partial r} \bfe_r + {1 \over r b^2} \bfb \otimes \frac{\partial \hat{\bfl}}{\partial \theta} \bfe_{\theta}
\end{align*}

Noting that $\hat{\bfl}=\bfe_{\theta}$ and hence, $\frac{\partial \hat{\bfl}}{\partial r}=0$ and $\frac{\partial \hat{\bfl}}{\partial \theta}=-\bfe_r$, we have 
\begin{align}\label{eq:divalpha}
div \bfalpha= {1 \over b^2} \bfb \otimes (0 . \bfe_r) - {1 \over r b^2}(\bfb \otimes \bfe_r ) \bfe_{\theta}= 0 - \frac{\bfb}{r b^2} \bfe_r \cdot \bfe_{\theta} = 0 + 0 =0. 
\end{align}


Also, 
\begin{align*}
{\bfalpha}:[{\bfalpha} ~grad\bfV]&= \frac{\bfb}{A}\otimes \hat{\bfl} : [(\frac{\bfb}{A}\otimes \hat{\bfl}) ~ grad\bfV] =  {1 \over A^2} ({\bfb}\otimes \hat{\bfl}) : \Big[{\bfb}\otimes [grad\bfV]^T ~ \hat{\bfl} \Big] \\
&= {1 \over b^4} ({\bfb}\cdot{\bfb}) ~ (\hat{\bfl} \cdot [grad\bfV]^T ~ \hat{\bfl}) 
\end{align*}
Since $\hat{\bfl}={\bfe}_{\theta}$, 
\begin{align*}
{\bfalpha}:[{\bfalpha} ~grad\bfV]&= {1 \over b^4} ({\bfb}\cdot{\bfb}) ~ \Big[{\bfe_{\theta}} \cdot \Big(\frac{\partial{V_r}}{\partial r} {\bfe_r} \otimes {\bfe_r} + \frac{V_r}{r} {\bfe_{\theta}} \otimes {\bfe_{\theta}}\Big)~ {\bfe_{\theta}} \Big] \\
&=  \frac{b^2}{b^4} ~ \Big[{\bfe_{\theta}} \cdot \frac{v(r)}{r} {\bfe}_{\theta} \Big] ={1 \over b^2} ~ \frac{v(r)}{r}
\end{align*}

Therefore, 
\begin{align}\label{eq:nonZeroTerm}
\overline{{\bfalpha}:[{\bfalpha}~ grad\bfV]}&= {1 \over \pi L^2 H} \int_{\theta=0}^{\theta=2\pi} \int_{r=0}^{r=L} \int_{z=-\frac{t}{2}}^{\frac{t}{2}} {1 \over b^2} ~ \frac{v}{r} r dr d\theta dz \nonumber \\
&=\frac{ t}{\pi L^2 H . b^2} \int_{\theta=0}^{\theta=2\pi} \int_{r=0}^{r=L}  \tilde{v} \{ H(r-R) - \tilde{v} H(r-(R+\Delta r)) \} dr d\theta \nonumber \\
&= \frac{t}{\pi L^2 H b^2 } \Big\{ \int_{r=R}^{r=R+\Delta R}  \tilde{v} dr \Big\} 2\pi = \frac{\tilde{v} t}{\pi L^2 H b^2} .  \Delta R . (2 \pi) \nonumber\\
&= \frac{2 \tilde{v} }{ L^2 H}
\end{align}

Substituting the results from \eqref{eq:vbar}, \eqref{eq:divVbar}, \eqref{eq:gradrho}, \eqref{eq:divalpha} and \eqref{eq:nonZeroTerm} in \eqref{eq:rhobardot_chap}, we get
\begin{align}\label{eq:rhobardot_loop}
\dot{\overline{\rho}}=& -\overline{0 \cdot {\bfV} } - \frac{2} {b^2} \overline{div V} + \overline{2 \bfalpha: (\bf0 \otimes \bfV)} + \frac{2 \tilde{v} }{ L^2 H} = 0 + 0 + 0 + \frac{2 \tilde{v} }{ L^2 H} \nonumber\\
=&  \frac{2 \tilde{v} }{ L^2 H}. 
\end{align} 

Since $\tilde{v} > 0$ for an expanding loop, this shows that $\dot{\overline{\rho}}>0$. This is justified because as shown before, $\bar{\rho}$ give the averaged line length and therefore it has to increase for an expanding loop. 

\subsubsection{The evolution equation for \textbf{plastic distortion rate}, $\bfL^p$} There are many quantities whose evolution are governed by the average of the fluctuation terms. For example, the evolution of ${\bfL}^p$ defined by \eqref{eq:Lp} and obtained using \eqref{eq:product_avg_genl} is 
\begin{equation}\label{lp_evol}
\begin{split}
\dot{{\bfL}^p}&= \dot{\overline{{\bfalpha}\times{\bfV}}} - \dot{\overline{\bfalpha}} \times \overline{\bfV} - \overline{\bfalpha} \times \dot{\overline{\bfV}} \\
&= \dot{\overline{\bfalpha}} \times \overline{\bfV} + \overline{\bfalpha} \times \dot{\overline{\bfV}} + \overline{ {\Sigma}^{\dot{\bfalpha}} \times {\Sigma}^{\bfV} +  {\Sigma}^{{\bfalpha}} \times {\Sigma}^{\dot{\bfV}} } - 
 \dot{\overline{\bfalpha}} \times \overline{\bfV} - \overline{\bfalpha} \times \dot{\overline{\bfV}} \\
&=\overline{ {\Sigma}^{-curl({\bfalpha}\times{\bfV})} \times {\Sigma}^{\bfV} +  {\Sigma}^{{\bfalpha}} \times {\Sigma}^{\dot{\bfV}} }.
\end{split}
\end{equation}
This shows that the evolution of $\bfL^p$ is governed by the sum of the averages of the fluctuation terms. 



For the example of an expanding circular loop, using the results from \eqref{eq:alphabar} and \eqref{eq:vbar} and the fact that $\hat{\bfl}=\bfe_{\theta}$, 
\begin{align}\label{eq:lp_loop}
\bfL^p=&\overline{\bfalpha \times \bfV} - \overline{\bfalpha} \times \overline{\bfV}=\overline{ (\frac{\bfb}{A} \otimes \hat{\bfl}) \times \bfV} - 0 \times 0 \nonumber \\
=& \overline{ \frac{v}{b^2} \bfb \otimes (\bfe_{\theta} \times \bfe_r)  }=\overline{-\frac{v}{b^2} \bfb \otimes \bfe_z} \nonumber \\
=& {1 \over \pi L^2 H} \int_{\theta=0}^{\theta=2\pi} \int_{r=0}^{r=L} \int_{z=-\frac{t}{2}}^{\frac{t}{2}} -\frac{v}{b^2} \bfb \otimes \bfe_z r dr d\theta dz \nonumber \\
=& -\frac{t} {\pi L^2 H. b^2} {\bfb} \otimes {\bfe_z} \int_{\theta=0}^{\theta=2\pi} \int_{r=0}^{r=L} v r dr d\theta \nonumber \\
=& -\frac{t} {\pi L^2 H. b^2} {\bfb} \otimes {\bfe_z} \int_{\theta=0}^{\theta=2\pi} \int_{r=0}^{r=L}  \tilde{v} \{ H(r-R) - \tilde{v} H(r-(R+\Delta r)) \} rdr d\theta \nonumber \\
=& -\frac{t} {\pi L^2 H. b^2} {\bfb} \otimes {\bfe_z}\Big\{ \int_{r=R}^{r=R+\Delta R}  \tilde{v} rdr \Big\} 2\pi \nonumber\\
=& -\frac{\tilde{v} t} {\pi L^2 H. b^2} {\bfb} \otimes {\bfe_z}  \left[ \frac{r^2}{2} \right]_R^{R+\Delta R} . 2\pi =-\frac{ 2 \tilde{v} t  \{ \Delta R . (2 R + \Delta R) \} } {\pi L^2 H. b^2} {\bfb} \otimes {\bfe_z}  \nonumber\\
\implies |\bfL^p|=& -\frac{ 2  \tilde{v} t . \Delta R . (2 R + \Delta R)} {L^2 H. b^2}. b= - \frac{ 2  \tilde{v} . (2 R + \Delta R).b} {L^2 H} , 
\end{align}
where $\bfe_z=\bfe_r \times \bfe_{\theta}$. From \eqref{eq:rhobardot_loop} and \eqref{eq:lp_loop}, we observe that both $\dot{\overline{\rho}}$ and $|\bfL^p|$, for the case of a uniformly expanding circular loop, are proportional to $\tilde{v}$, and hence, $\dot{\overline{\rho}}$ is proportional to $|\bfL^p|$.  This observation is in agreement with classical theory which states that $\dot{\bar{\rho}}$ (where the averaged line length $\overline{\rho}$ is a measure of the strength of the material) is proportional to $|\bfL^p|$. However, in classical theory, strength of a material cannot decrease whereas $\bar{\rho}$ can decrease in our case.

\subsection{Crystal Plasticity MFDM}\label{subsec:refined_desc}

Conventional crystal plasticity involves resolution of the system of evolution equations into individual slip systems and superposing the effect of plastic strain on different slip systems. Motivated by the work in \cite{acharya_chapman} to evaluate what is involved in working with the evolution of slip-system level coarse variables (as proposed in \cite{hochrainer_materials, xia_elazab}, but using ad-hoc equations of mesoscopic evolution as discussed in Section \ref{sec:intro}), we consider a refined description, in which we define state variables with respect to individual slip system and derive their evolution. The state variables that describe this model are:
\begin{subequations} \label{eq:refined}
\begin{align}
& \quad \bfalpha, & \nonumber \\
& \quad {\bfa}^l : = \chi^l {\bfalpha}, & \label{eq:def_a_l}  \\
& \quad \rho^l : ={\bfa}^l:{\bfa}^l. & \label{eq:def_rho_l}
\end{align}
\end{subequations}
Here, ${\bfalpha}$ is the dislocation density tensor, $\chi^l ({\bfx},t)$ is the characteristic function of dislocations of slip system $l$ (with normal ${\bfn}^l$ and slip direction ${\bfb}^l$) at position $\bfx$ and $\bfa^l$ and $\rho^l$ are the dislocation density tensor and total dislocation density respectively, corresponding to slip system $l$. 

The characteristic function $\chi^l({\bfx},t)$ indicates whether the point $\bfx$ at time $t$ is occupied by a dislocation of slip system $l$. We denote the exponential operator as $e(.)$. The characteristic function can be approximated as
\begin{align}\label{eq:chi_l}
\chi^l({\bfx},t) \approx e \left( - {\left(\frac{|{\bfalpha} {\bfn}^l |} {c_1} \right)}^m \right) ~ e\left(-{\left( \frac{||{\tilde{\bfb}	}^l.\tilde{\bfalpha} \tilde{\bfalpha}^T . \tilde{\bfb}^l| -1|} {c_2} \right)}^n \right),
\end{align}
where 
\begin{align}
\tilde{\bfb}^l=\frac{\bfb^l}{|{\bfb^l}|} \nonumber\\
\tilde{\bfalpha}=\frac{\bfalpha}{|{\bfalpha}|}
\end{align} 
and $c_1$ and $c_2$ are very small positive constants and $m$ and $n$ are very large positive constants. For a dislocation to belong to slip system $l$, it must satisfy $\bfalpha \cdot \bfn^l=0$ (as $\bfalpha \cdot \bfn^l=(\bfb \otimes \hat{\bfl}) \cdot \bfn^l=(\hat{\bfl} \cdot \bfn^l)\bfb= 0.\bfb=0 $, where $\bfb$ and $\hat{\bfl}$ are the Burgers vector and line direction of the dislocation respectively). In that case, the first term on the rhs of \eqref{eq:chi_l}, $e \left( - {\left(\frac{|{\bfalpha} {\bfn}^l |} {c_1} \right)}^m \right)$, is 1 as $\frac{|{\bfalpha} {\bfn}^l |} {c_1}=0$. Otherwise (when $\bfalpha \cdot \bfn^l \neq 0$), the first term $e \left( - {\left(\frac{|{\bfalpha} {\bfn}^l |} {c_1} \right)}^m \right)$ is approximately 0, since $c_1$ is a small positive constant and $m$ is a very large positive constant.
Moreover, if the Burgers vector $\bfb$ of the dislocation coincides with $\bfb^l$, the term $|{\tilde{\bfb}	}^l.\tilde{\bfalpha} \tilde{\bfalpha}^T . \tilde{\bfb}^l| =1$. Hence, in that case, the second term on the rhs of \eqref{eq:chi_l} is 1, since $\left( \frac{||{\tilde{\bfb}	}^l.\tilde{\bfalpha} \tilde{\bfalpha}^T . \tilde{\bfb}^l| -1|} {c_2} \right)=0$. Otherwise, it is approximately 0, since $c_2$ is a small positive constant and $n$ is a very large positive constant. Thus, the first term decides whether the dislocation is in the slip plane of the slip system, while the second term decides whether it has the same Burgers vector as the slip system. Only when both of these are true, we have $\chi^l =1$. Otherwise, we have $\chi^l \approx 0$.

An implied assumption in the definition of the characteristic function and the slip system variables is that a particular spatial location is occupied at any instant by a dislocation of a single slip system, which excludes the proper accounting of junctions in the definition of the slip system variables, even though the microscopic dynamics does not involve any such exclusion.

\subsection{The coarse variables} \label{sec:coarse_vars}
We are interested in understanding the plastic behavior of metals at a length scale which is much coarser than the atomic length scale and at a time scale comparable to real life load applications (which is much larger compared to the time scale of the motion of dislocations, set by the drag). Therefore, we are interested in the averaged theory of the microscopic dynamics, which involves the evolution of the coarse variables corresponding to the variables defined in \eqref{eq:refined}, which are $\overline{\bfalpha}$, $\overline{\bfa^l}$ and $\overline{\rho^l}$. In order to do so, we also need to know the evolution of the averaged characteristic function $\overline{\bfchi^l}$, since $\chi^l$ appears on the rhs of their evolution equations \eqref{eq:refined}. In this Section, we derive the evolution of these coarse variables. Corresponding brackets have been marked with the same color to make the equations look more readable. The primary averaged variables have been marked in blue, to show how much of the rhs is known in terms of them.  We also define the following variables to make the equations look more readable and compact: 
\begin{align*}
P&:=\frac{|{\bfalpha} {\bfn}^l |} {c_1} \\
Q&:= \frac{||{\tilde{\bfb} }^l.\tilde{\bfalpha} \tilde{\bfalpha}^T. \tilde{\bfb}^l|-1|} {c_2} \\ 
p&:={\tilde{\bfb} }^l.\tilde{\bfalpha} \tilde{\bfalpha}^T. \tilde{\bfb}^l. 
\end{align*}
By their definition, $P, Q$, and $p$ are indexed by the slip-system indicator $l$, and this will be understood in the following without explicit notation.
\subsubsection{The evolution equation for averaged dislocation density, $\overline{\bfalpha}$}
The evolution of $\overline{\bfalpha}$ (following \cite{acharya_roy_2006}) is given by 
\begin{align}\label{eq:alphabar}
\dot{{\color{blue}\overline{\bfalpha}}}=-curl({\color{blue}\overline{\bfalpha}} \times {\color{blue}\overline{\bfV}} + {\bfL}^p). 
\end{align}

\subsubsection{The evolution equation for the averaged characteristic function, $\overline{\chi^l}$, for slip system $l$}
$\overline{\chi^l}$ is obtained by applying the averaging operator \eqref{eq:avg} to \eqref{eq:chi_l}. Its evolution equation is  
\begin{align}\label{eq:chibardot_main}
{\color{blue}\dot{\overline{\chi^l}}}=: &  \ \mathfrak{B}^l(state) = -\frac{m}{c_1} ~\overline{\ePm} ~ \overline{\eQn} ~ \overline{\PmMinusOne} ~ \Big[ - {\color{blue}\overline{ {\bfalpha}} }{\bfn}^l  \cdot \Big\{{curl( {\color{blue}\overline{\bfalpha}} \times {\color{blue}\overline{\bfV}} + {\bfL}^p) {\bfn}^l} . \overline{\left(1 \over {|{\bfalpha} {\bfn}^l}| \right)} \Big\} \Big] \nonumber\\
&- \frac{n}{c_2}  ~\overline{\ePm} ~ \overline{\eQn} ~\overline{\QnMinusOne} ~\overline{sgn(|p|-1)} ~\overline{ sgn(p) } ~   \nonumber\\
&\quad  \tilde{{\bfb}}^l \cdot {\color{brown}\Bigg\{ } {\color{magenta}\Bigg(} {-curl( {\color{blue}\overline{\bfalpha}} \times {\color{blue}\overline{\bfV}}+ {\bfL}^p )}  ~\overline{ \Big({1 \over |\bfalpha|}\Big)}  +{\color{blue}\overline{\bfalpha}} : {curl( {\color{blue}\overline{\bfalpha}} \times {\color{blue}\overline{\bfV}} + {\bfL}^p )} ~ \overline{\left(1 \over {|\bfalpha|^3}\right)} {\color{magenta}\Bigg)} ~ \overline{\tilde{\bfalpha}^T} \nonumber\\
& \quad \qquad + {\overline{\tilde{\bfalpha}}}  {\color{cyan}\Bigg( } ({-curl({\color{blue}\overline{\bfalpha}} \times {\color{blue}\overline{\bfV}} + {\bfL}^p )})^T ~ \overline{ \Big({1 \over |\bfalpha|}\Big)} +{\color{blue}\overline{\bfalpha}} : {curl({\color{blue}\overline{\bfalpha}} \times {\color{blue}\overline{\bfV}} + {\bfL}^p )}   \overline{\Big({1 \over {|\bfalpha|^3}}\Big)} {\color{cyan}\Bigg)} {\color{brown}\Bigg\}} \cdot \tilde{\bfb^l}\nonumber\\
& - \frac{m}{c_1} {\color{blue}\Bigg\{ } ~\overline{\ePm} ~ \overline{\eQn} ~ \overline{\PmMinusOne} ~ {\color{brown}\Bigg(} \overline{ {\Sigma}^{{\bfalpha} {\bfn}^l} \cdot {\Sigma}^{{ -curl({\bfalpha} \times {\bfV}) {\bfn}^l  }} } \overline{\left( 1 \over {|{\bfalpha} {\bfn}^l}|\right)}   \nonumber\\
& \qquad \qquad \qquad \qquad \qquad \qquad \qquad + \overline{ {\Sigma}^{-({\bfalpha} {\bfn}^l) \cdot { (curl({\bfalpha} \times {\bfV}) {\bfn}^l)}}   ~ {{\Sigma}^{1 \over {|{\bfalpha} {\bfn}^l}|}  } } {\color{brown}\Bigg)} \nonumber\\
& \quad \qquad + \overline{ {\Sigma}^{{\ePm} ~ {\eQn}} }  ~ \overline{\PmMinusOne} ~ {\color{cyan} \Bigg(} {\PbarDotFirst}\nonumber \\
& \quad \qquad + {\PbarDotSecond} {\color{cyan}\Bigg)} \nonumber\\
&\quad \qquad + \overline{ {\Sigma}^{ {\ePm} ~ {\eQn}}  ~ {\Sigma}^{\PmMinusOne} } ~{\color{magenta}\Bigg(} {\PbarDotFirst} \nonumber \\
& \quad \qquad + {\PbarDotSecond} {\color{magenta}\Bigg)}  {\color{blue}\Bigg\} } \nonumber\\
&- \frac{n}{c_2} {\color{blue}\Bigg[ } \overline{\ePm} ~ \overline{\eQn} ~ \overline{\QnMinusOne} \nonumber\\
& \quad \qquad { \overline{sgn(|p|-1)} ~\overline{ sgn(p) } ~ \tilde{{\bfb}}^l \cdot {\color{brown}\Bigg\{ } {\color{cyan}\Bigg( }  \overline{ {\Sigma}^{-curl(\bfalpha \times \bfV)} {\Sigma}^{1 \over {|\bfalpha|} } } } \nonumber\\
&\quad \qquad -{\Big[ \Big\{ \Big( \overline{{\Sigma}^{\bfalpha} : {\Sigma}^{-curl({\bfalpha}\times {\bfV})}} \Big) \overline{ \Big({1 \over {|\bfalpha|^3}}\Big)}} \nonumber\\
&\quad \qquad +{\QbarDotThree} \nonumber\\
&\quad \qquad +{ {\overline{\tilde{\bfalpha}}} ~ {\color{magenta}\Bigg(} \overline{{\Sigma}^{-(curl(\bfalpha \times \bfV))^T} {\Sigma}^{1 \over {|\bfalpha|} } }} \nonumber\\
&\quad \qquad \qquad \quad -{\Big[ \Big\{ \Big( \overline{{\Sigma}^{\bfalpha} : {\Sigma}^{-(curl({\bfalpha}\times {\bfV}))^T}} \Big) \overline{ \Big({1 \over {|\bfalpha|^3}}\Big)}}  \nonumber\\ 
&\quad \qquad \qquad \quad +{\QbarDotSix}  \nonumber\\  
&\quad \qquad +{\QbarDotSeven}  \nonumber\\ 
&\quad \qquad {\QbarDotEight}  \nonumber\\  
&\quad \qquad +{\QbarDotNine}  \nonumber\\  
&\quad \qquad {\QbarDotTen}  \nonumber\\  
& \quad \qquad + \overline{ {\Sigma}^{{\ePm} ~ {\eQn}} }  ~ \overline{\QnMinusOne} \nonumber\\
& \quad \quad \qquad {\QbarDotOne} \nonumber\\
&\quad \quad \qquad -{\QbarDotTwo} \nonumber\\
&\quad \quad  \qquad +{\QbarDotThree} \nonumber\\
&\quad \quad  \qquad +{\QbarDotFour} \nonumber\\
&\quad \quad  \qquad \qquad \quad -{\QbarDotFive}  \nonumber\\ 
&\quad \quad  \qquad \qquad \quad +{\QbarDotSix}  \nonumber\\ 
&\quad \quad \qquad +{\QbarDotSeven}  \nonumber\\  
&\quad \quad \qquad {\QbarDotEight}  \nonumber\\ 
&\quad \quad \qquad +{\QbarDotNine}  \nonumber\\  
&\quad \qquad {\QbarDotTen}  \nonumber\\  
& \quad \qquad + \overline{ {\Sigma}^{{\ePm} ~ {\eQn}} ~ {\Sigma}^{\QnMinusOne} }  \nonumber \\
&\quad \quad \qquad {\QbarDotOne} \nonumber\\
&\quad \quad \qquad -{\QbarDotTwo} \nonumber\\
&\quad \quad \qquad +{\QbarDotThree} \nonumber\\
&\quad \quad \qquad +{\QbarDotFour} \nonumber \\
&\quad \quad \qquad \qquad \quad -{\QbarDotFive} \nonumber \\
&\quad \quad \qquad \qquad \quad +{\QbarDotSix} \nonumber \\
&\quad \quad \qquad +{\QbarDotSeven}  \nonumber\\ 
&\quad \quad \qquad {\QbarDotEight}  \nonumber\\  
&\quad \quad \qquad +{\QbarDotNine}  \nonumber\\  
&\quad \quad \qquad {\QbarDotTen} {\color{blue} \Bigg] }  \nonumber\\  
&- m~\overline{ {\Sigma}^{{\ePm} ~ {\eQn} ~ {\PmMinusOne} }  ~ {\Sigma}^{\Pdot} } \nonumber \\
&- n~ \overline{ \, {\color{blue}\Bigg(} {\Sigma}^{{\ePm} ~ {\eQn} ~ {\QnMinusOne} } } \nonumber\\ 
& \quad \qquad \overline{ {\Sigma}^{ \frac{1}{c_2}{\QdotOne} + {\QdotTwo}}   {\color{blue}\Bigg)} \, },
\end{align}
where $\mathfrak{B}^l(state)$ represents the state function given by the rhs of \eqref{eq:chibardot_main}. 
The derivation of \eqref{eq:chibardot_main} is given in Appendix \ref{app:chi_l}. 

%

\subsubsection{The evolution equation for the averaged dislocation density tensor, $\overline{\bfa^l}$, for slip system $l$}\label{sec:evol_al_bar}
$\overline{\bfa^l}$ is obtained by applying the averaging operator \eqref{eq:avg} to \eqref{eq:def_a_l}. Its evolution equation is 
\begin{align} \label{eq:bfabarl_evol_main}
{\color{blue}\dot{\overline{\bfa^l}}} &=\mathfrak{B}^l(state) ~ {\color{blue}\overline{\bfalpha}}  - curl[ {\color{blue}\overline{{\chi}^l}} \, ( {\color{blue}\overline{\bfa^l}} \times  {\color{blue} \overline{\bfV}} ) ] - curl ({\color{blue}\overline{\bfa^l}} \times \overline{\Sigma^{\chi^l} \Sigma^{\bfV} })  + 2 ~ {\color{blue}\overline{\chi^l}} ~ ({\color{blue}\overline{\bfalpha}} \times {\color{blue}\overline{\bfV}}) [{\bfX}(grad {\color{blue}\overline{\chi^l}})] \nonumber\\
& \quad + \overline{{\Sigma}^{\dot{\chi}} ~ {\Sigma}^{\bfalpha}}  - curl\Big(\overline{ {\Sigma}^{\bfa^l} \times {\Sigma}^{{\bfV}^l}}\Big) + 2 ~ {\color{blue}\overline{\chi^l}} ~\overline{ {\Sigma}^{\bfalpha} \times {\Sigma}^{\bfV} } [{\bfX}(grad {\color{blue}\overline{\chi^l}})] + 2 ~ \overline { {\Sigma}^{\chi^l} ~ {\Sigma}^{{\bfalpha} \times {\bfV} } } [{\bfX}(grad {\color{blue}\overline{\chi^l}})] \nonumber \\
& \quad +2~ \overline{ {\Sigma}^{\chi^l ({\bfalpha}\times {\bfV}) } ~ {\Sigma}^{{\bfX}(grad_{x'} \chi^l)} },
\end{align}
where $\mathfrak{B}^l$ is defined in the discussion following \eqref{eq:chibardot_main} in Section \ref{sec:coarse_vars}. 
The derivation of \eqref{eq:bfabarl_evol_main} is given in Appendix \ref{app:a_l}.  The merit of \eqref{eq:bfabarl_evol_main} is that it shows what the exact evolution equation of $\overline{{\bfa}^l}$ should be (cf. \cite{xia_elazab}). It is cumbersome, to say the least and, moreover, contains fluctuation terms whose evolution are given by other pde, resulting in an `unsolvable' infinite hierarchy. 

%

%

\subsubsection{The evolution equation for the averaged total dislocation density, $\overline{\rho^l}$, for slip system $l$} 
$\overline{\rho^l}$ is obtained by applying the averaging operator \eqref{eq:avg} to \eqref{eq:def_rho_l}. Its evolution is given by 
\begin{align}\label{eq:rho_l_dot_bar_main}
{\color{blue}\dot{\overline{\rho^l}}}=&-2\, \mathfrak{B}^l(state)\, {\color{blue}\overline{\rho^l}} - grad{{\color{blue}\overline{\rho^l}}} \cdot ({\color{blue}\overline{\chi^l} \, \overline{\bfV}}) + {\color{blue}\overline{\rho}} ~ grad{{\color{blue}\overline{\chi^l}} } \cdot ({\color{blue}{\overline{\chi^l}} \, \overline{V}})  - 2 ~ {\color{blue}\overline{\rho^l}} ~ div({\color{blue}{\overline{\chi^l} \,\overline{\bfV}}}) + 2 ~ {\color{blue}\overline{\rho^l}} ~ grad{{\color{blue}\overline{\chi^l}}} \cdot {\color{blue}{\overline{\bfV}}} \nonumber \\
& \qquad  +  2 ~ ( {\color{blue}\overline{\chi^l} ~\overline{\bfalpha}}) : ( {\color{blue}\overline{\bfalpha}} grad {\color{blue}\overline{\bfV}} )  +  2 ~ ( {\color{blue}\overline{\chi^l} ~\overline{\bfalpha}}) : ( div {\color{blue}\overline{\bfalpha}} \otimes {\color{blue}\overline{\bfV}} ) \nonumber \\  
&+ 2 ~ \overline{{\Sigma}^{{\color{red}\Bigg[}-m~ \ePm \eQn \PmMinusOne \Pdot - \frac{n}{c_2} ~\ePm \eQn \QnMinusOne \cdot } } \nonumber\\
& \qquad \, \overline{ {}^{  {\color{magenta}\Bigg(} \QdotOne + \QdotTwo {\color{magenta} \Bigg)} {\color{red}\Bigg]}}  {\Sigma}^{\rho^l}} \nonumber\\
& - grad{{\color{blue}\overline{\rho^l}}} \cdot \overline{\Sigma^{\chi^l}\, \Sigma^V} + {\color{blue}\overline{\rho}} ~ grad{{\color{blue}\overline{\chi^l}} } \cdot \overline{\Sigma^{\chi^l}\, \Sigma^V}  - 2 ~ {\color{blue}\overline{\rho^l}} ~ div(\overline{\Sigma^{\chi^l}\, \Sigma^V}) \nonumber\\
& - \overline{ {\Sigma}^{grad \rho^l} \cdot {\Sigma}^{\chi^l \bfV} } \qquad + \overline{ {\Sigma}^{\rho}~{\Sigma}^{grad {\chi^l}} } \cdot ( {\color{blue}\overline{\chi^l} \, \overline{V}} + \overline{{\Sigma}^{\chi^l} \Sigma^{\bfV}}) + \overline{ {\Sigma}^{\rho ~ grad{\chi^l}} \cdot {\Sigma}^{\chi^l \, \bfV}}   - 2 ~ \overline{ {\Sigma}^{\rho^l} ~ {\Sigma}^{div (\chi^l \, \bfV)} }  \nonumber \\
&  + 2 ~ \overline{ {\Sigma}^{\rho^l} ~ {\Sigma}^{grad{\chi^l} } } \cdot {\color{blue}\overline{\bfV}} + 2 ~ \overline{ {\Sigma}^{\rho^l ~ grad{\chi^l}} \cdot {\Sigma}^{\bfV} } + \overline{ {\Sigma}^{\chi^l} ~ {\Sigma}^{\bfalpha} }: \Big({\color{blue}\overline{ {\bfalpha}}} ~ grad {\color{blue}\overline{\bfV}} + \overline{ {\Sigma}^{\bfalpha}~{\Sigma}^{grad{\bfV}} } \Big) +\overline{ {\Sigma}^{{\chi^l} {\bfalpha}}: {\Sigma}^{{\bfalpha}\, grad \bfV }} \nonumber \\
& + 2 ~ \overline{ {\Sigma}^{\chi^l} ~ {\Sigma}^{\bfalpha} }: ( div {\color{blue}\overline{\bfalpha}} \otimes {\color{blue}\overline{\bfV}} + \overline{ {\Sigma}^{div \bfalpha} \otimes {\Sigma}^{\bfV}}  ) + \overline{ {\Sigma}^{{\chi^l} {\bfalpha}}: {\Sigma}^{div {\bfalpha} \otimes \bfV}} \nonumber \\
\end{align}
where $\mathfrak{B}^l(state)$ is defined in the discussion following \eqref{eq:chibardot_main} in Section \ref{sec:coarse_vars}. The derivation of \eqref{eq:rho_l_dot_bar_main} is given in Appendix \ref{app:rho_l}. The equation \eqref{eq:rho_l_dot_bar_main} is the exact evolution equation of $\overline{{\rho}^l}$. The same remarks as to the practicality of  this exact equation as in Section \ref{sec:evol_al_bar} applies. 

\textbf{Remark} 
In the evolution equations for $\bar{\bfalpha}$ \eqref{eq:alphabar}, $\bar{\bfa}^l$ \eqref{eq:bfabarl_evol_main} and $\bar{\rho}^l$ \eqref{eq:rho_l_dot_bar_main}, the plastic distortion rate $\bfL^p$ appears. It is defined in \eqref{eq:Lp} and is a fluctuation term ($\bfL^p=\overline{\Sigma^{\bfalpha} \times \Sigma^{\bfV}}$). As shown in Section \ref{sec:hierarchy}, the hierarchy can involve equations of evolution for the fluctuations. We derived the evolution equation for $\bfL^p$ in \eqref{lp_evol} which is as follows: 
\[
\dot{{\bfL}^p}=\overline{ {\Sigma}^{-curl({\bfalpha}\times{\bfV})} \times {\Sigma}^{\bfV} +  {\Sigma}^{{\bfalpha}} \times {\Sigma}^{\dot{\bfV}}}. 
\]
The $\bfL^p$ for a uniformly expanding circular loop was obtained in Section \ref{sec:iso_mfdm} and is given by \eqref{eq:lp_loop}. However, this was possible due to the drastic assumption of uniform velocity (of same magnitude pointing radially outward) at all points of the loop. In reality, the value of the local velocity is difficult to obtain without consideration of the microscopic DD problem, as it depends on the Peach-Koehler force acting on the dislocation segments, which is a function of the internal stresses. This makes it essentially impossible to define an evolution equation for $\bfL^p$ in realistic situations without some sort of `on-the-fly' coupling to local DD calculations. The coupled DD-MFDM strategy that is described and implemented in \cite{ddfdm_coupling} defines evolution equations for $\bfL^p$ using appropriate time averaging of Discrete Dislocation Dynamics is a first demonstration towards achieving exactly this goal for realistic applied loading rates. 

\section{Conclusion}
We stated some descriptors of the microscopic dynamics and obtained the evolution of the coarse variables generated from such descriptors. The coarse variables give an idea of the averaged behavior of the system at a much coarser length and time scale. We see that the evolution of the total dislocation density \eqref{eq:rhobardot} contains the averages of fluctuations, and hence is exact but not closed. We considered a refined description in which we resolved the dynamics into slip systems. We see that the evolution of the dislocation density tensor \eqref{eq:bfabarl_evol_main} and the total dislocation density \eqref{eq:rho_l_dot_bar_main} of any particular slip system is extremely cumbersome, which shows the limitations of such a refined description. The evolution equations of the coarse variables involve many average terms, average of fluctuation terms and their partial derivatives, all of which have their own evolution given by other pdes. Thus, we get an infinite hierarchy of non-linear non-closed coarse evolution pdes, which cannot be solved for all practical purposes. The CDD framework \cite{hochrainer2007three, hochrainer2016thermodynamically, sandfeld2015pattern} \textit{postulates} the microscopic dynamics and uses closure assumptions of their own to cut off the infinite hierarchy of equations. In contrast, MFDM \eqref{eq:MFDM}, which follows by averaging the equations of FDM \eqref{eq:FDM}  in space and time, is based on the fundamental statement of the conservation of Burgers vector (which is a physically observed fact). While cumbersome, one could try to work with these exact equations if they were known in full detail. If this is not the case, the justifications for using such infinite hierarchies is scarce. It is much more reasonable, and important to focus on closure assumptions, generated from the actual stress coupled microscopic dislocation interaction dynamics and their averaging, at a lower level in the hierarchy, about which such `kinematic' infinite hierarchies say nothing. 

In previous work starting from \cite{acharya_roy_2006}, the system is closed using physics-based phenomenological modeling at the lowest level of the hierarchy as a trade-off with practicality (see the discussion surrounding \eqref{eq:MFDM}, in which $\bfL^p$ and $\bfV$ are phenomenologically specified). This approach has been quite successful in addressing a promising array of problems in modern plasticity theory related to the computation of patterning, dislocation internal stress, size effects, polygonization, and slip transmission at grain boundaries among others \cite{roy2006size,acharya2007jump,puri2010modeling,
mach2010continuity, fressengeas2011dislocation,puri2011controlling,
puri2011mechanical,acharya2011microcanonical, das2016microstructure, fressengeas2009dislocation, taupin2007effects, taupin2010particle,richeton2011continuity,
taupin2008directionality,djaka2015numerical,varadhan2009lattice,
djaka2020fft,berbenni2020fast, genee2020particle}, including long-standing and recent fundamental challenges in the prediction of large-deformation, dislocation mediated elastic and elastic-plastic response \cite{arora2020dislocation, arora2020finite,
arora2020unification}. Despite the phenomenology, the approach has provided two distinct benefits:
\begin{enumerate}
\item It is fair to say that beyond the modeling in 1 space dimension, the plastic distortion in classical plasticity theory has, at best, only a thermodynamic physical meaning with no concrete, tangible connection to the mechanics of dislocations. The MFDM approach based on space-time averaging of microscopic dislocation mechanics brings out an explicit, completely defined, connection between the plastic strain rate employed in phenomenological theories of plasticity and the motion and geometry of an evolving microscopic array of dislocations, as explained in the discussion surrounding \eqref{eq:MFDM}. Obviously, this has many benefits, even for a phenomenological specification of the macroscopic plastic strain rate. Moreover, the MFDM framework has allowed for a first unification between phenomenological $J_2$ and crystal plasticity theories and quantitative dislocation mechanics. 
\item With a single extra material parameter beyond a classical plasticity model (and two at finite deformations), the MFDM framework has enabled a significant variety of phenomena to be modeled, in qualitative and quantitative accord with experimental results. 
\end{enumerate}

In the work presented in \cite{ddfdm_coupling}, the phenomenological constitutive assumptions in MFDM are replaced with inputs obtained by appropriate time averaging of Discrete Dislocation Dynamics. To our knowledge, this is the first demonstration of a coupled DD-continuum plasticity framework that can be exercised at quasi-static loading rates and makes no assumptions on material response beyond the elasticity of the material and a model for thermal activation of dislocations past sessile junctions (which is not a part of the microscopic DD model). For reasons mentioned in \cite{ddfdm_coupling}, a first systematic improvement of the DD-MFDM coupled model would be to add the equation of the evolution of the total density \eqref{eq:rhobardot} to the field equations of MFDM (with its non-closed terms involving fluctuations supplied by local space-time averaged DD response), which would provide a further local driving constraint (the value of $\bar{\rho}$) to each of the local DD calculations beyond the stress of the macroscopic model. Of course, even with dropping all the fluctuation terms in \eqref{eq:rhobardot}, that equation can augment the phenomenological mode of application of MFDM, enhancing the description of material strength beyond the currently prevalent non-decreasing phenomenological descriptions (e.g. the Voce Law). These are some of the practical benefits of pursuing the space-time averaged hierarchy of dislocation mechanics as developed in this paper.

\begin{appendices}
\section{Derivation of evolution equations}
\subsection{Total dislocation density, $\rho^l$}\label{subsec:rhobardot}
From \eqref{eq:rho}, we have
\begin{align}\label{eq:rho_app}
\rho={\bfalpha}:{\bfalpha}
\end{align}

We differentiate \eqref{eq:rho_app} in time to get
\begin{align}\label{eq:rhodot}
\dot{\rho}&=2 ~ {\bfalpha}:\dot{\bfalpha}=-2 ~ {\bfalpha}:curl({\bfalpha}\times {\bfV}) =-2 \alpha_{ij} [curl({\bfalpha} \times {\bfV})]_{ij} \nonumber \\
&=-2 \alpha_{ij} e_{jmn} ({\bfalpha}\times{\bfV})_{in,m} = -2 \alpha_{ij} e_{jmn} e_{npq} (\alpha_{ip} V_q)_{,m} \nonumber \\
&=-2 (\delta_{jp} \delta_{mq} - \delta_{jq} \delta_{mp}) [\alpha_{ij} \alpha_{ip,m} V_q + \alpha_{ij} \alpha_{ip} V_{q,m}] \nonumber \\
&=-2 [\alpha_{ij} \alpha_{ij,m} V_m + \alpha_{ij} \alpha_{ij} V_{m,m} - \alpha_{ij} \alpha_{im,m} V_j - \alpha_{ij} \alpha_{im} V_{j,m}] \nonumber \\
&=-2 ~[ {1 \over 2} grad({\bfalpha}:{\bfalpha}) \cdot \bfV + {\bfalpha}:{\bfalpha} (div V) - \bfalpha: (div \bfalpha \otimes \bfV) - {\bfalpha}: \{{\bfalpha} \,grad V\} ] \nonumber \\
&=-2 ~[ {1 \over 2} grad \rho \cdot {\bfV} + \rho (div V) - \bfalpha: (div \bfalpha \otimes \bfV) - {\bfalpha}: \{{\bfalpha} \,grad V\} ] \nonumber \\
&= -grad \rho \cdot {\bfV} - 2 \rho (div V) +2 \bfalpha: (div \bfalpha \otimes \bfV) + 2 {\bfalpha}: \{{\bfalpha} \,grad V\} 
\end{align}
We apply the averaging operator to both sides of \eqref{eq:rhodot}  to get 
\begin{align}\label{eq:rhodot2}
\dot{\overline{\rho}} = - grad~ \overline{\rho} \cdot \overline{\bfV} - 2 ~\overline{\rho}~ div \overline{\bfV}  - \overline{ {\Sigma}^{grad {\rho}} \cdot {\Sigma}^{\bfV}} - 2 \overline{{\Sigma}^{\rho} {\Sigma}^{div V}} + 2 ~\overline{ \bfalpha: (div \bfalpha \otimes \bfV)  }  + 2 ~ \overline{ {\bfalpha}: \{ {\bfalpha} ~ grad {\bfV}\} }.
\end{align}
Using \eqref{eq:rhodot2} and the facts that
\begin{align*}
\overline{ {\bfalpha}: \{ {\bfalpha} ~grad {\bfV}\} }&= \overline{\bfalpha}: \{ \overline{\bfalpha} ~ grad \overline{\bfV} \} + \overline{\bfalpha} : \overline{ {\Sigma}^{\bfalpha} ~ {\Sigma}^{grad{\bfV}} } + \overline{ {\Sigma}^{\bfalpha} :  {\Sigma}^{ {\bfalpha} ~ {grad{\bfV}} }  } \\
\overline{ \bfalpha: (div \bfalpha \otimes \bfV)  } &= \overline{\bfalpha} : ( div \overline{\bfalpha} \otimes \overline{\bfV} ) + \overline{\bfalpha} : ( \overline{ \Sigma^{div \bfalpha}  \otimes \Sigma^{\bfV}} ) + \overline{ \Sigma^{\bfalpha} : \Sigma^{ div \bfalpha \otimes \bfV} },
\end{align*}
we get the evolution of $\overline{\rho}$ as \eqref{eq:rhobardot} in section \ref{sec:conn_with_mfdm}. 

\subsection{The characteristic function, $\chi^l$}\label{app:chi_l}

We will use the following results in the derivation: 
\begin{itemize}
\item
If $\bff$ is a vector, then 
\begin{align}\label{eq:vecModDot}
\frac{d}{dt} {|\bff|}=\frac{ {\bff} \cdot \dot{\bff} } {|\bff|}, 
\end{align}
which gives, 
\begin{align}\label{eq:vecModBarDot}
\dot{\overline{|{\bff}|}}=\frac{\overline{{\bff} \cdot \dot{\bff}}} {|{\bff}|} = \overline{ {\bff} \cdot {\dot{\bff}}} . \overline{ 1 \over |{\bff}|}  + \overline{ {\Sigma}^{ {\bff} \cdot \dot{\bff} } {\Sigma}^{ 1 \over {|{\bff}|}}} = \left(\overline{{\bff}} \cdot \dot{ \overline{{\bff}} } + \overline{ {\Sigma}^{\bff} \cdot {\Sigma}^{\dot{\bff}} } \right) . \overline{1 \over {|\bff|}} + \overline{ {\Sigma}^{ {\bff} \cdot \dot{\bff} } {\Sigma}^{ 1 \over {|{\bff}|}}}.
\end{align}

\item 
If $q$ is a scalar, 
\begin{align}\label{eq:scaModDot}
\frac{d}{dt} {|q|}=sgn(q) \dot{q},
\end{align}
which gives,
\begin{align}\label{eq:scaModBarDot}
\dot{\overline{|q|}} = \overline{ sgn(q) ~ \dot{q} } = \overline{sgn(q)} ~ \dot{\overline{q}} + \overline{ {\Sigma}^{sgn(q)} {\Sigma}^{\dot{q}}}.
\end{align}
\end{itemize}

We define 
\begin{align}\label{eq:chi}
\chi^l:= \ePm ~ \eQn, 
\end{align}
where $P=\frac{|{\bfalpha} {\bfn}^l |} {c_1} $ and $Q= \frac{||{\tilde{\bfb} }^l.\tilde{\bfalpha} \tilde{\bfalpha}^T  \tilde{\bfb}^l|-1|} {c_2} $. We also denote $p={\tilde{\bfb} }^l.\tilde{\bfalpha} \tilde{\bfalpha}^T  \tilde{\bfb}^l$. Hence, $Q=\frac{||p|-1|}{c_2}$. Taking time derivative of \eqref{eq:chi} and using \eqref{eq:vecModDot} and \eqref{eq:scaModDot}, we have, 
\begin{equation}\label{eq:chidot}
\begin{split}
&\dot{\chi^l}=-m ~ \ePm \eQn {P}^{m-1} \dot{P} - n ~ \ePm \eQn {Q}^{n-1} \dot{Q} 
\end{split}
\end{equation}
From \eqref{eq:chidot}, we have, 
\begin{equation}\label{eq:chibar_l_dot}
\begin{split}
\dot{\overline{\chi^l}}=& -m \Big[ ~ \overline{\ePm ~ \eQn ~ {P}^{m-1}} ~\dot{\overline{P}} + \overline{ {\Sigma}^{\ePm ~ \eQn ~ {P}^{m-1} } ~ {\Sigma}^{\dot{P}} } \Big] \\
&- n ~ \Big[\overline{\ePm ~ \eQn ~ {Q}^{n-1}} ~\dot{\overline{Q}} + \overline{{\Sigma}^{\ePm~ \eQn ~ {Q}^{n-1}} ~ {\Sigma}^{\dot{Q}} }  \Big] \\
= & - m ~ \overline{\ePm ~ \eQn ~ {P}^{m-1}} ~\dot{\overline{P}}  - n ~ \overline{\ePm ~ \eQn ~ {Q}^{n-1}} ~\dot{\overline{Q}} \\
&-m ~ \overline{ {\Sigma}^{\ePm ~ \eQn ~ {P}^{m-1} } ~ {\Sigma}^{\dot{P}} } - n ~ \overline{{\Sigma}^{\ePm ~ \eQn ~ {Q}^{n-1}} ~ {\Sigma}^{\dot{Q}} }  \\
= & -m \Big(~ \overline{\ePm ~ \eQn} ~ \overline{{P}^{m-1}} + \overline{ {\Sigma}^{\ePm ~ \eQn} ~ {\Sigma}^{P^{m-1}} } \Big) ~\dot{\overline{P}} \\
&- n \Big(~ \overline{\ePm ~ \eQn} ~ \overline{{Q}^{n-1}} + \overline{ {\Sigma}^{\ePm ~ \eQn} ~ {\Sigma}^{Q^{n-1}} } \Big) ~\dot{\overline{Q}}  \\
& - m ~ \overline{ {\Sigma}^{\ePm ~ \eQn ~ {P}^{m-1} } ~ {\Sigma}^{\dot{P}} } - n ~ \overline{{\Sigma}^{\ePm ~ \eQn ~ {Q}^{n-1}} ~ {\Sigma}^{\dot{Q}} }  \\
= & -m \Big(~ ( \overline{\ePm} ~ \overline{\eQn} + \overline{ {\Sigma}^{\ePm ~ \eQn} } ) ~ \overline{{P}^{m-1}} + \overline{ {\Sigma}^{\ePm ~ \eQn} ~ {\Sigma}^{P^{m-1}} } \Big) ~\dot{\overline{P}} \\
&- n \Big(~ ( \overline{\ePm} ~ \overline{\eQn} + \overline{ {\Sigma}^{\ePm ~\eQn} } ) ~ \overline{{Q}^{n-1}} + \overline{ {\Sigma}^{\ePm ~ \eQn} ~ {\Sigma}^{Q^{n-1}} } \Big) ~\dot{\overline{Q}}  \\
&- m ~ \overline{ {\Sigma}^{\ePm ~ \eQn ~ {P}^{m-1} } ~ {\Sigma}^{\dot{P}} } - n ~ \overline{{\Sigma}^{\ePm ~ \eQn ~ {Q}^{n-1}} ~ {\Sigma}^{\dot{Q}} }  \\
= &-m \Big( ~ \overline{\ePm} ~ \overline{\eQn} ~ \overline{{P}^{m-1}} ~\dot{\overline{P}} + \overline{ {\Sigma}^{\ePm ~ \eQn }}  ~ \overline{{P}^{m-1}} ~\dot{\overline{P}} + \overline{ {\Sigma}^{\ePm ~ \eQn} ~ {\Sigma}^{P^{m-1}} }  ~\dot{\overline{P}} \Big) \\
&- n \Big( ~ \overline{\ePm} ~ \overline{\eQn} ~ \overline{{Q}^{n-1}} ~\dot{\overline{Q}} + \overline{ {\Sigma}^{\ePm ~ \eQn} }  ~ \overline{{Q}^{n-1}} ~\dot{\overline{Q}} + \overline{ {\Sigma}^{\ePm ~ \eQn} ~ {\Sigma}^{Q^{n-1}} }  ~\dot{\overline{Q}} \Big)  \\
&- m~\overline{ {\Sigma}^{\ePm ~ \eQn ~ {P}^{m-1} } ~ {\Sigma}^{\dot{P}} } -n~ \overline{{\Sigma}^{\ePm ~ \eQn ~ {Q}^{n-1}} ~ {\Sigma}^{\dot{Q}} }
\end{split}
\end{equation}

From the definition of $P$ in the discussion following \eqref{eq:chi} and using \eqref{eq:vecModDot}, 
\begin{align}
\dot{P}= {1 \over c_1}  \frac {({\bfalpha} ~ {\bfn}^l) \cdot (\dot{{\bfalpha}} ~ {\bfn}^l)} { |{\bfalpha} ~ {\bfn}^l |}=  -{1 \over c_1}  \frac {({\bfalpha} ~ {\bfn}^l) \cdot \{curl({\bfalpha} \times {\bfV}) ~ {\bfn}^l \} } { |{\bfalpha} ~ {\bfn}^l |}
\end{align}

Using \eqref{eq:vecModBarDot}, 
\begin{align}\label{eq:T1dot}
\dot{\overline{P}}&={1 \over {c_1}} \left[ \left( \overline{ {\bfalpha} {\bfn}^l} \cdot \dot{\overline{{\bfalpha} {\bfn}^l}}  + \overline{{\Sigma}^{{\bfalpha} {\bfn}^l} \cdot {\Sigma}^{{ \dot{\bfalpha} {\bfn}^l  }} }\right) \overline{ \left(  1 \over {|{\bfalpha} {\bfn}^l}| \right) }  + \overline{ {\Sigma}^{({\bfalpha} {\bfn}^l) \cdot { (\dot{\bfalpha}} {\bfn}^l)}   ~ {\Sigma}^{1 \over {|{\bfalpha} {\bfn}^l}|}  } \right] \nonumber \\
&={1 \over {c_1}} \left[ \left( (\overline{ {\bfalpha}} {\bfn}^l) \cdot ({\dot{\overline{\bfalpha}} {\bfn}^l})  + \overline{{\Sigma}^{{\bfalpha} {\bfn}^l} \cdot {\Sigma}^{{ \dot{\bfalpha} {\bfn}^l  }} } \right) \overline{ \left(  1 \over {|{\bfalpha} {\bfn}^l}| \right)}  + \overline{ {\Sigma}^{({\bfalpha} {\bfn}^l) \cdot { (\dot{\bfalpha}} {\bfn}^l)}   ~ {\Sigma}^{1 \over {|{\bfalpha} {\bfn}^l}|}  } \right] \nonumber \\
&={1 \over {c_1}} \Big[  \Big( -(\overline{ {\bfalpha}} {\bfn}^l) \cdot ({curl( \overline{\bfalpha} \times \overline{\bfV} + {\bfL}^p) {\bfn}^l})  +  \overline{{\Sigma}^{{\bfalpha} {\bfn}^l} \cdot {\Sigma}^{{ \dot{\bfalpha} {\bfn}^l  }} }\Big) \overline{\Big(  {1 \over {|{\bfalpha} {\bfn}^l}| } \Big) }  \nonumber \\
&\qquad + \overline{ {\Sigma}^{-({\bfalpha} {\bfn}^l) \cdot { (curl({\bfalpha} \times {\bfV}) {\bfn}^l)}}   ~ {{\Sigma}^{1 \over {|{\bfalpha} {\bfn}^l}|}  } } \Big].
\end{align}

From the definition of $p$ and $Q$ in the discussion following \eqref{eq:chi} and using \eqref{eq:scaModDot},
\begin{align}\label{eq:Qdot}
\dot{Q}={1 \over c_2} sgn(|p|-1) ~ sgn(p)~\dot{p}={1 \over c_2} sgn(|p|-1) ~ sgn(p)  ~ \tilde{{\bfb}}^l \cdot ({ \dot{\tilde{\bfalpha}} \tilde{\bfalpha}^T + \tilde{\bfalpha} \dot{\tilde{\bfalpha}}^T) \tilde{{\bfb}}^l}   
\end{align}
Now, 
\begin{align}\label{eq:alphatildedot}
\dot{\tilde{\bfalpha}}&=\frac{\dot{\bfalpha}}{|\bfalpha|} - \left( \frac{ {\bfalpha}:\dot{\bfalpha} }{ |\bfalpha|^3} \right) ~ \bfalpha = -\frac{curl({\bfalpha}\times {\bfV})}{|\bfalpha|} + \left( \frac{ {\bfalpha}:curl({\bfalpha}\times{\bfV}) }{ |\bfalpha|^3} \right) ~ \bfalpha \nonumber \\
\dot{\widetilde{\bfalpha}^T}&= \frac{\dot{\bfalpha}^T}{|\bfalpha|}= -\frac{\{curl({\bfalpha}\times{\bfV})\}^T}{|\bfalpha|} + \left( \frac{ {\bfalpha}:\{curl({\bfalpha}\times{\bfV})\} }{ |\bfalpha|^3} \right) ~ {\bfalpha}^T
\end{align}
Substituting \eqref{eq:alphatildedot} into \eqref{eq:Qdot}, 
\begin{align}\label{eq:Qdotfinal}
\dot{Q}&={1 \over c_2} sgn(|p|-1) ~ sgn(p)  ~ \tilde{{\bfb}}^l \cdot \Big[ \Big\{ -\frac{curl({\bfalpha}\times {\bfV})}{|\bfalpha|} + \big( \frac{ {\bfalpha}:curl({\bfalpha}\times{\bfV}) }{ |\bfalpha|^3} \big)  ~ \bfalpha \Big\} \tilde{\bfalpha}^T \nonumber \\
&+ \tilde{\bfalpha} \Big\{ -\frac{\{curl({\bfalpha}\times{\bfV})\}^T}{|\bfalpha|}  + \big( \frac{ {\bfalpha}:\{curl({\bfalpha}\times{\bfV})\} }{ |\bfalpha|^3} \big)  ~ {\bfalpha}^T \Big\} \Big] \tilde{{\bfb}}^l   
\end{align}

Using \eqref{eq:Qdot}, 
\begin{equation}\label{eq:Qbardot}
\begin{split}
\dot{\overline{Q}}&={1 \over c_2} \overline{ sgn(|p|-1) ~ sgn(p) ~ \dot{p} } = {1 \over c_2} \Big[ \overline{sgn(|p|-1)} ~ \overline{sgn(p) ~ \dot{p}} + \overline{{\Sigma}^{sgn(|p|-1)} {\Sigma}^{sgn(p) \dot{p}}} \Big] \\
& = {1 \over c_2} \Big[ \overline{sgn(|p|-1)} \left\{ \overline{sgn(p)} ~ \dot{\overline{p}} + \overline{ {\Sigma}^{sgn(p)} {\Sigma}^{\dot{p}}} \right\} + \overline{{\Sigma}^{sgn(|p|-1)} {\Sigma}^{sgn(p) \dot{p}}} \Big] \\
& = {1 \over c_2} \Big[ \overline{sgn(|p|-1)} ~ \overline{sgn(p)} ~ \dot{\overline{p}} + \overline{sgn(|p|-1)} ~ \overline{ {\Sigma}^{sgn(p)} {\Sigma}^{\dot{p}}} + \overline{{\Sigma}^{sgn(|p|-1)} {\Sigma}^{sgn(p) \dot{p}}} \Big]
\end{split}
\end{equation}

Using the definition of $p$ in the discussion around \eqref{eq:Qdot} ,
\begin{align}\label{eq:pdot}
\dot{p}=\tilde{{\bfb}}^l \cdot ({ \dot{\tilde{\bfalpha}} \tilde{\bfalpha}^T + \tilde{\bfalpha} \dot{\tilde{\bfalpha}}^T) \tilde{{\bfb}}^l}  \nonumber \\
\implies \dot{\overline{p}}= \tilde{{\bfb}}^l \cdot{ (\dot{\overline{\tilde{\bfalpha}}} ~ \overline{\tilde{\bfalpha}^T} + {\overline{\tilde{\bfalpha}}} ~\dot{ \overline{{{\widetilde{\bfalpha}^T}}}}) \tilde{{\bfb}}^l}
\end{align}

Following \eqref{eq:MFDM}, we have 
\begin{align*}
\dot{\overline{\bfalpha}}=-curl(\overline{\bfalpha} \times \overline{\bfV} + {\bfL}^p)
\end{align*}
From \eqref{eq:alphatildedot},
\begin{align}\label{eq:alphadot}
\dot{\overline{\tilde{\bfalpha}} }&= \left( \dot{\overline{\bfalpha}} ~\overline{ 1 \over |\bfalpha|} + \overline{ {\Sigma}^{\dot{\bfalpha}} {\Sigma}^{1 \over {|\bfalpha|} } } \right) - \left\{  \overline{ \left( \frac{ {\bfalpha}:\dot{\bfalpha} }{ |\bfalpha|^3} \right)} \, \overline{\bfalpha} + \overline{ {\Sigma}^{\frac{ {\bfalpha}:\dot{\bfalpha} }{ |\bfalpha|^3} } {\Sigma}^{\bfalpha} } \right\}  \nonumber \\
&= \left( \dot{\overline{\bfalpha}} ~ \overline{ 1 \over |\bfalpha|} + \overline{ {\Sigma}^{\dot{\bfalpha}} {\Sigma}^{1 \over {|\bfalpha|} } } \right) - \left\{  \overline{ \left( \frac{ {\bfalpha}:\dot{\bfalpha} }{ |\bfalpha|^3} \right)}  \overline{\bfalpha} + \overline{ {\Sigma}^{\frac{ {\bfalpha}:\dot{\bfalpha} }{ |\bfalpha|^3} } {\Sigma}^{\bfalpha} } \right\}   \nonumber \\
&=\left( \dot{\overline{\bfalpha}}  ~\overline{ 1 \over |\bfalpha|} + \overline{ {\Sigma}^{\dot{\bfalpha}} {\Sigma}^{1 \over {|\bfalpha|} } } \right) - \left[ \left\{ \left( \overline{\bfalpha} : \dot{\overline{\bfalpha}} + \overline{{\Sigma}^{\bfalpha} : {\Sigma}^{\dot{\bfalpha}}} \right) \overline{1 \over {|\bfalpha|^3}} + \overline{{\Sigma}^{ {\bfalpha}:\dot{\bfalpha}} {\Sigma}^{1 \over {|\bfalpha|^3}} } \right\} \overline{\bfalpha} + \overline{ {\Sigma}^{\frac{ {\bfalpha}:\dot{\bfalpha} }{ |\bfalpha|^3} } {\Sigma}^{\bfalpha} } \right]   \nonumber \\
&=\Big( {-curl(\overline{\bfalpha} \times \overline{\bfV} + {\bfL}^p )}  ~\overline{ 1 \over |\bfalpha|} + \overline{ {\Sigma}^{-curl(\bfalpha \times \bfV)} {\Sigma}^{1 \over {|\bfalpha|} } } \Big)  
- \Bigg[ {\color{brown}\Bigg\{ } {\color{magenta}\Big(} -\overline{\bfalpha} : {curl(\overline{\bfalpha} \times \overline{\bfV} + {\bfL}^p )} \nonumber \\
&+ \overline{{\Sigma}^{\bfalpha} : {\Sigma}^{-curl({\bfalpha}\times {\bfV})}} {\color{magenta}\Big)} \overline{1 \over {|\bfalpha|^3}}
+ \overline{{\Sigma}^{ {\bfalpha}:{-curl({\bfalpha} \times {\bfV})}} {\Sigma}^{1 \over {|\bfalpha|^3}} } {\color{brown}\Bigg\}} \overline{\bfalpha} 
+ \overline{ {\Sigma}^{\frac{ -{\bfalpha}:curl({\bfalpha} \times {\bfV}) }{ |\bfalpha|^3} } {\Sigma}^{\bfalpha} } \Bigg].
\end{align}

Similarly, we can obtain $\dot{\overline{\widetilde{\bfalpha}^T}}=\frac{\bfalpha^T}{|\bfalpha|}$ by replacing $\bfalpha$, $\dot{\bfalpha}$, $\overline{\bfalpha}$ and $\dot{\overline{\bfalpha}}$ above with their respective transpose and obtain
\begin{align}\label{eq:alphaTdot}
\dot{\overline{\widetilde{\bfalpha}^T} }&=\Big( ({-curl(\overline{\bfalpha} \times \overline{\bfV} + {\bfL}^p )})^T ~ \overline{ 1 \over |\bfalpha|} + \overline{ {\Sigma}^{-(curl(\bfalpha \times \bfV))^T} {\Sigma}^{1 \over {|\bfalpha|} } } \Big)    
- \Bigg[ {\color{brown}\Bigg\{} {\color{magenta}\Big(} -\overline{\bfalpha} : { (curl(\overline{\bfalpha} \times \overline{\bfV} + {\bfL}^p )}) \nonumber\\
&+ \overline{{\Sigma}^{\bfalpha^T} : {\Sigma}^{-(curl({\bfalpha} \times {\bfV}))^T}} {\color{magenta}\Big)} \overline{1 \over {|\bfalpha|^3}}  
+ \overline{{\Sigma}^{ -{\bfalpha}:{(curl({\bfalpha} \times {\bfV}))}} {\Sigma}^{1 \over {|\bfalpha|^3}} } {\color{brown}\Bigg\}} \overline{\bfalpha^T} + \overline{ {\Sigma}^{\frac{ -{\bfalpha}:(curl({\bfalpha} \times {\bfV}))}{ |\bfalpha|^3} } {\Sigma}^{\bfalpha^T} } \Bigg].
\end{align}

Using \eqref{eq:Qbardot}, \eqref{eq:pdot}, \eqref{eq:alphadot} and \eqref{eq:alphaTdot}, we get
\begingroup
\allowdisplaybreaks
\begin{align}
\dot{\overline{Q}}&={1 \over c_2} \QbarDotOne \nonumber \\
&\qquad \qquad - \QbarDotTwo \nonumber \\
& \qquad \qquad \qquad + \QbarDotThree \nonumber \\
& \qquad \qquad + \QbarDotFour \nonumber \\
& \qquad \qquad \qquad \quad - \QbarDotFive \nonumber \\
& \qquad \qquad \qquad \quad + \QbarDotSix \nonumber \\
&+ {1 \over c_2} \QbarDotSeven \nonumber \\
& \QbarDotEight \nonumber \\
&+ {1 \over c_2} \QbarDotNine \nonumber \\
& \QbarDotTen 
\end{align}
\endgroup

Using \eqref{eq:chidot}, \eqref{eq:T1dot}, \eqref{eq:Qdot}, \eqref{eq:pdot}, \eqref{eq:alphadot} and \eqref{eq:alphaTdot}, we get the evolution of $\overline{\chi^l}$ given by \eqref{eq:chibardot_main} in section \ref{sec:conn_with_mfdm}.

\subsection{Dislocation density tensor corresponding to slip system $l$, $\bfa^l$}\label{app:a_l}
Following \eqref{eq:refined}, the dislocation density corresponding to slip system $l$ is defined as
\begin{align}\label{eq:refined_app}
\bfa^l:= \chi^l \bfalpha.
\end{align}

We take time derivative of \eqref{eq:refined_app} to obtain
\begin{align}\label{eq:bfadot}
\dot{\bfa^l}&=\dot{\chi^l}{\bfalpha} + \chi^l \dot{{\bfalpha}} \nonumber \\
		&=\dot{\chi^l}{\bfalpha} - \chi^l curl({{\bfalpha} \times {\bfV}}).\nonumber \\
\end{align}

Using the fact that $\chi^l \approx ({\chi^l})^2$, since $\chi^l$ can (approximately) take either of the values 0 or 1, the second term on the right hand side above can be written as
\begin{align}\label{eq:second_term}
&\chi^l curl({\bfalpha} \times {\bfV})= \chi^l \{curl({\bfalpha} \times {\bfV})\}_{im} \approx ({\chi^l})^2 e_{mjk} \{{\bfalpha} \times {\bfV}\}_{ik,j} \nonumber \\
&=e_{mjk} \{ \chi^l {\bfalpha} \times {\chi^l} {\bfV} \}_{ik,j} -   e_{mjk} \{{\bfalpha} \times {\bfV}\}_{ik}  \frac{\partial {\chi^l}^2}{\partial {\chi_{j'}}}\nonumber \\
&=\{curl( \chi^l {{\bfalpha} \times  \chi^l {\bf V}})\}_{im} - 2\{{\bfalpha} \times {\bfV}\}_{ik} e_{mjk} \chi^l \frac{\partial \chi^l}{\partial \chi_{j'}} \nonumber \\
&=\{curl( \bfa^l \times {\bfV}^l)\}_{im} - 2 ~ \chi^l ~\{{\bfalpha} \times {\bfV}\}_{ik} [{\bfX}(grad_{x'} \chi^l)]_{km} \nonumber \\
&\Rightarrow \chi^l curl({\bfalpha} \times {\bfV}) = curl( \bfa^l \times {\bfV}^l) -  2 ~ \chi^l ({\bfalpha} \times {\bfV}) [{\bfX}(grad_{x'} \chi^l)],  
\end{align} 
where $e_{mjk}$ is a component of the third-order alternating tensor $\bfX$ and its action on a tensor $\bfA$ is given by $\{{\bfX}({\bfA})\}_i=e_{ijk}A_{jk} $, while its action on a vector $\bfN$ is given by $\{{{\bfX}({\bfN})}\}_{ij}=e_{ijk}N_{k} $. 

Using \eqref{eq:bfadot} and \eqref{eq:second_term} above, we have
\begin{align}\label{eq:bfadot_final}
\dot{\bfa^l}=\dot{\chi^l} {\bfalpha} - curl(\bfa^l \times {\bfV}^l) + 2 ~ \chi^l ({\bfalpha} \times {\bfV}) [{\bfX}(grad_{x'} \chi^l)].
\end{align}

We average both sides of \eqref{eq:bfadot_final} to get
\begin{align}\label{eq:bfabardot}
\dot{\overline{\bfa^l}}&=\dot{\overline{\chi^l}} ~\overline{\bfalpha} + \overline{{\Sigma}^{\dot{\chi}^l} ~ {{\Sigma}^{\bfalpha}}} - curl(\overline{\bfa^l} \times \overline{{\bfV}^l}) - curl\Big(\overline{ {\Sigma}^{\bfa^l} \times {\Sigma}^{{\bfV}^l}}\Big) +  \overline{2 \chi^l ({\bfalpha}\times {\bfV})[{\bfX}(grad_{x'} \chi^l)]} 
\end{align}

We have
\begin{align*}
&\overline{2 \chi^l ({\bfalpha}\times {\bfV})[{\bfX}(grad_{x'} \chi^l)]}= 2 ~ \overline{\chi^l ({\bfalpha}\times {\bfV})} ~ [{\bfX}(grad \overline{\chi^l})] + 2 ~ \overline{ {\Sigma}^{\chi^l ({\bfalpha}\times {\bfV}) } ~ {\Sigma}^{{\bfX}(grad_{x'} \chi^l)} } \\
&=2 ~ \Big(  \overline{\chi^l} ~ \overline{{\bfalpha} \times {\bfV}}   + \overline { {\Sigma}^{\chi^l} ~ {\Sigma}^{{\bfalpha} \times {\bfV} } } \Big) ~ [{\bfX}(grad \overline{\chi^l})] + 2 ~ \overline{ {\Sigma}^{\chi^l ({\bfalpha}\times {\bfV}) } ~ {\Sigma}^{{\bfX}(grad_{x'} \chi^l)} } \\
&= 2 ~ \overline{\chi^l} ~ (\overline{\bfalpha} \times \overline{\bfV}) [{\bfX}(grad \overline{\chi^l})] + 2 ~ \overline{\chi^l} ~\overline{ {\Sigma}^{\bfalpha} \times {\Sigma}^{\bfV} } [{\bfX}(grad \overline{\chi^l})] + 2 ~ \overline { {\Sigma}^{\chi^l} ~ {\Sigma}^{{\bfalpha} \times {\bfV} } } [{\bfX}(grad \overline{\chi^l})] \\
& \quad +2~ \overline{ {\Sigma}^{\chi^l ({\bfalpha}\times {\bfV}) } ~ {\Sigma}^{{\bfX}(grad_{x'} \chi^l)} }
\end{align*}

Hence, using \eqref{eq:bfabardot}, we have 
\begin{align} \label{eq:bfabardot_final}
&\dot{\overline{\bfa^l}}=\dot{\overline{\chi^l}} ~\overline{\bfalpha} + \overline{{\Sigma}^{\dot{\chi}} ~ {\Sigma}^{\bfalpha}} - curl(\overline{\bfa^l} \times \overline{{\bfV}^l}) - curl\Big(\overline{ {\Sigma}^{\bfa^l} \times {\Sigma}^{{\bfV}^l}}\Big) \nonumber\\
&+ 2 ~ \overline{\chi^l} ~ (\overline{\bfalpha} \times \overline{\bfV}) [{\bfX}(grad \overline{\chi^l})] + 2 ~ \overline{\chi^l} ~\overline{ {\Sigma}^{\bfalpha} \times {\Sigma}^{\bfV} } [{\bfX}(grad \overline{\chi^l})] + 2 ~ \overline { {\Sigma}^{\chi^l} ~ {\Sigma}^{{\bfalpha} \times {\bfV} } } [{\bfX}(grad \overline{\chi^l})] \nonumber \\
& \quad +2~ \overline{ {\Sigma}^{\chi^l ({\bfalpha}\times {\bfV}) } ~ {\Sigma}^{{\bfX}(grad_{x'} \chi^l)} }
\end{align}

Finally, we use \eqref{eq:chidot} and \eqref{eq:chibardot_main} to obtain the evolution of $\overline{\bfa^l}$ as \eqref{eq:bfabarl_evol_main} in section \ref{sec:conn_with_mfdm}. 

\subsection{Total dislocation density corresponding to slip system $l$, $\rho^l$}\label{app:rho_l}\label{app:rho_l}

The total dislocation density corresponding to slip system $l$ is defined in \eqref{eq:refined} and is given by 
\begin{align}\label{eq:rho_l}
\rho^l:= {\bfa^l}:{\bfa^l} = ({\chi^l \, \bfalpha}) : ({\chi^l \, \bfalpha}) \approx ({\chi^l})^2 ~ {\bfalpha}:{\bfalpha}.
\end{align}

We differentiate \eqref{eq:rho_l} with respect to time to get 
\begin{align*}
\dot{\rho^l} &=2 {\chi^l} ~ \dot{\chi^l} ~ {\bfalpha}:{\bfalpha} + 2 ~({\chi^l})^2 ~ {\bfalpha}:\dot{\bfalpha} \\
&= 2 \dot{\chi^l} {\rho^l} + {\chi^l} [2 {\bfalpha} : \dot{\bfalpha}].
\end{align*}

Using \eqref{eq:rhodot}, we have, 
\begin{align} \label{eq:rholDot}
\dot{\rho^l} &= 2 \dot{\chi^l} {\rho^l} - 2 {\chi^l} \Big[ {1 \over 2} grad \rho \cdot {\bfV} + \rho (div V) - {\bfalpha}: \{{\bfalpha} \,grad V\}  - \bfalpha: (div \bfalpha \otimes \bfV) \Big] \nonumber \\
& \approx 2 \dot{\chi^l} {\rho^l} -  ({\chi^l})^2 grad\rho \cdot {\bfV} -2 ({\chi^l})^2 \rho ~div\bfV + 2 {\chi^l} {\bfalpha} : ({\bfalpha}\, grad \bfV) + 2 {\chi^l} {\bfalpha} : (div \bfalpha \otimes \bfV)  \nonumber \\
&= 2 \dot{\chi^l} {\rho^l} -  \{ {\chi^l} grad\rho \} \cdot ({\chi^l} {\bfV}) -2 ({\chi^l} \rho) \{{\chi^l} ~ div\bfV\} + 2 {\chi^l} {\bfalpha} : ({\bfalpha}\, grad \bfV) + 2 {\chi^l} {\bfalpha} : (div \bfalpha \otimes \bfV) \nonumber \\
&= 2 \dot{\chi^l} {\rho^l} -  \{ grad\rho^l - \rho ~ grad\chi^l \} \cdot {\bfV^l} -2  {\rho^l} \{ div{\bfV}^l - grad \chi^l \cdot {\bfV} \} + 2 {\chi^l} {\bfalpha} : ({\bfalpha}\, grad \bfV) + 2 {\chi^l} {\bfalpha} : (div \bfalpha \otimes \bfV) \nonumber \\
&= 2 \dot{\chi^l} {\rho^l} -  grad\rho^l \cdot {\bfV^l}  + \rho ~ grad\chi^l \cdot {\bfV^l} -2  {\rho^l} div{\bfV}^l + 2 {\rho^l} grad \chi^l \cdot {\bfV} \nonumber\\
& \quad + 2 {\chi^l} {\bfalpha} : ({\bfalpha}\, grad \bfV) + 2 {\chi^l} {\bfalpha} : (div \bfalpha \otimes \bfV). 
\end{align}

We average both sides of \eqref{eq:rholDot} to get the evolution of $\overline{\rho^l}$ as 
\begin{align}
&\dot{\overline{\rho^l}}= 2 ~ \dot{\overline{\chi^l}} ~ \overline{\rho^l} + 2 \overline{ {\Sigma}^{\dot{\chi^l}} ~ {\Sigma}^{\rho^l} } - grad{\overline{\rho^l}} \cdot \overline{\bfV^l} - \overline{ {\Sigma}^{grad \rho^l} \cdot {\Sigma}^{\bfV^l} } + \overline{\rho ~grad{\chi^l} } \cdot {\overline{\bfV^l}} + \overline{ {\Sigma}^{\rho ~ grad{\chi^l}} \cdot {\Sigma}^{\bfV^l}}  \nonumber\\
& \quad - 2 ~ \overline{\rho^l} ~ div{\overline{\bfV^l}} - 2 ~ \overline{ {\Sigma}^{\rho^l} ~ {\Sigma}^{div {\bfV^l}} } + 2 ~ \overline{{\rho^l} ~ grad{\chi^l}} \cdot {\overline{\bfV}} + 2 ~ \overline{ {\Sigma}^{\rho^l ~ grad{\chi^l}} \cdot {\Sigma}^{\bfV} }  \nonumber\\
& \quad + 2 ~ \overline{{\chi^l} {\bfalpha} : ({\bfalpha}\, grad \bfV)} + 2 ~ \overline{{\chi^l} {\bfalpha} : (div {\bfalpha} \otimes \bfV)} 
\end{align}

Also,
\begin{align*}
&\overline{{\chi^l} {\bfalpha} : ({\bfalpha}\, grad \bfV)}= \overline{ {\chi^l} {\bfalpha} } : \overline{{\bfalpha}\, grad \bfV } + \overline{ {\Sigma}^{{\chi^l} {\bfalpha}}: {\Sigma}^{{\bfalpha}\, grad \bfV }} \nonumber\\
&= ( \overline{\chi^l} ~\overline{\bfalpha} + \overline{ {\Sigma}^{\chi^l} ~ {\Sigma}^{\bfalpha} }): (\overline{ {\bfalpha}} ~ grad \overline{\bfV} + \overline{ {\Sigma}^{\bfalpha}~{\Sigma}^{grad{\bfV}} } ) + \overline{ {\Sigma}^{{\chi^l} {\bfalpha}}: {\Sigma}^{{\bfalpha}\, grad \bfV }}
\end{align*}

Similarly,
\begin{align*}
&\overline{{\chi^l} {\bfalpha} : (div {\bfalpha} \otimes \bfV)} = \overline{ {\chi^l} {\bfalpha} } : \overline{div {\bfalpha} \otimes \bfV } + \overline{ {\Sigma}^{{\chi^l} {\bfalpha}}: {\Sigma}^{div {\bfalpha} \otimes \bfV}} \nonumber\\
&= ( \overline{\chi^l} ~\overline{\bfalpha} + \overline{ {\Sigma}^{\chi^l} ~ {\Sigma}^{\bfalpha} }): ( div \overline{\bfalpha} \otimes \overline{\bfV} + \overline{ {\Sigma}^{div \bfalpha} \otimes {\Sigma}^{\bfV}}  ) + \overline{ {\Sigma}^{{\chi^l} {\bfalpha}}: {\Sigma}^{div {\bfalpha} \otimes \bfV}}
\end{align*}

Hence, 
\begin{equation}
\begin{split}
\dot{\overline{\rho^l}} &= 2 ~ \dot{\overline{\chi^l}} ~ \overline{\rho^l} + 2 \overline{ {\Sigma}^{\dot{\chi^l}} ~ {\Sigma}^{\rho^l} } - grad{\overline{\rho^l}} \cdot \overline{\bfV^l} - \overline{ {\Sigma}^{grad \rho^l} \cdot {\Sigma}^{\bfV^l} }  + \overline{\rho} ~ grad{\overline{\chi^l} } \cdot {\overline{\bfV^l}} + \overline{ {\Sigma}^{\rho}~{\Sigma}^{grad {\chi^l}} } \cdot \overline{\bfV^l}   \\
& + \overline{ {\Sigma}^{\rho ~ grad{\chi^l}} \cdot {\Sigma}^{\bfV^l}}   - 2 ~ \overline{\rho^l} ~ div{\overline{\bfV^l}}   - 2 ~ \overline{ {\Sigma}^{\rho^l} ~ {\Sigma}^{div {\bfV^l}} } + 2 ~ \overline{\rho^l} ~ grad{\overline{\chi^l}} \cdot {\overline{\bfV}} + 2 ~ \overline{ {\Sigma}^{\rho^l} ~ {\Sigma}^{grad{\chi^l} } } \cdot \overline{\bfV} \\
& \quad + 2 ~ \overline{ {\Sigma}^{\rho^l ~ grad{\chi^l}} \cdot {\Sigma}^{\bfV} } + \Big( \overline{\chi^l} ~\overline{\bfalpha} + \overline{ {\Sigma}^{\chi^l} ~ {\Sigma}^{\bfalpha} }\Big): \Big(\overline{ {\bfalpha}} ~ grad \overline{\bfV} + \overline{ {\Sigma}^{\bfalpha}~{\Sigma}^{grad{\bfV}} } \Big) +\overline{ {\Sigma}^{{\chi^l} {\bfalpha}}: {\Sigma}^{{\bfalpha}\, grad \bfV }} \\
& \quad + 2 ~ ( \overline{\chi^l} ~\overline{\bfalpha} + \overline{ {\Sigma}^{\chi^l} ~ {\Sigma}^{\bfalpha} }): ( div \overline{\bfalpha} \otimes \overline{\bfV} + \overline{ {\Sigma}^{div \bfalpha} \otimes {\Sigma}^{\bfV}}  ) + \overline{ {\Sigma}^{{\chi^l} {\bfalpha}}: {\Sigma}^{div {\bfalpha} \otimes \bfV}} \\
 \implies \dot{\overline{\rho^l}} &= 2 ~ \dot{\overline{\chi^l}} ~ \overline{\rho^l} - grad{\overline{\rho^l}} \cdot \overline{\bfV^l} + \overline{\rho} ~ grad{\overline{\chi^l} } \cdot {\overline{\bfV^l}}  - 2 ~ \overline{\rho^l} ~ div{\overline{\bfV^l}} + 2 ~ \overline{\rho^l} ~ grad{\overline{\chi^l}} \cdot {\overline{\bfV}} \\
& \quad  +  2 ~ ( \overline{\chi^l} ~\overline{\bfalpha}) : ( \overline{\bfalpha} grad \overline{\bfV} )  +  2 ~ ( \overline{\chi^l} ~\overline{\bfalpha}) : ( div \overline{\bfalpha} \otimes \overline{\bfV} ) + 2 \overline{ {\Sigma}^{\dot{\chi^l}} ~ {\Sigma}^{\rho^l} } - \overline{ {\Sigma}^{grad \rho^l} \cdot {\Sigma}^{\bfV^l} } \\
& \qquad + \overline{ {\Sigma}^{\rho}~{\Sigma}^{grad {\chi^l}} } \cdot \overline{\bfV^l} + \overline{ {\Sigma}^{\rho ~ grad{\chi^l}} \cdot {\Sigma}^{\bfV^l}}   - 2 ~ \overline{ {\Sigma}^{\rho^l} ~ {\Sigma}^{div {\bfV^l}} } + 2 ~ \overline{ {\Sigma}^{\rho^l} ~ {\Sigma}^{grad{\chi^l} } } \cdot \overline{\bfV} \\
& \quad + 2 ~ \overline{ {\Sigma}^{\rho^l ~ grad{\chi^l}} \cdot {\Sigma}^{\bfV} } + \overline{ {\Sigma}^{\chi^l} ~ {\Sigma}^{\bfalpha} }: \Big(\overline{ {\bfalpha}} ~ grad \overline{\bfV} + \overline{ {\Sigma}^{\bfalpha}~{\Sigma}^{grad{\bfV}} } \Big) +\overline{ {\Sigma}^{{\chi^l} {\bfalpha}}: {\Sigma}^{{\bfalpha}\, grad \bfV }} \\
& \quad + 2 ~ \overline{ {\Sigma}^{\chi^l} ~ {\Sigma}^{\bfalpha} }: ( div \overline{\bfalpha} \otimes \overline{\bfV} + \overline{ {\Sigma}^{div \bfalpha} \otimes {\Sigma}^{\bfV}}  ) + \overline{ {\Sigma}^{{\chi^l} {\bfalpha}}: {\Sigma}^{div {\bfalpha} \otimes \bfV}} \\
\end{split}
\end{equation}

Finally, using \eqref{eq:chidot} and \eqref{eq:chibardot_main}, we have the evolution equation for $\overline{\rho^l}$ as \eqref{eq:rho_l_dot_bar_main} in section \ref{sec:conn_with_mfdm}.

\end{appendices}

\section*{Acknowledgment}
Support from NSF grant NSF-CMMI-1435624 is gratefully acknowledged.

\bibliographystyle{Bibliography_Style}
\bibliography{testbibliog}

\end{document}

%% file: latexConfigureFile.tex
\usepackage{amsmath}
\usepackage{amsbsy}
\usepackage{amssymb}
\usepackage{amscd}
\usepackage{amsfonts}

\usepackage{isomath}
\usepackage{amsmath}
\usepackage{amsbsy}
\usepackage{amssymb}
\usepackage{amscd}
\usepackage{amsfonts}
\usepackage{graphicx}
\usepackage{verbatim}
\usepackage{euscript}
\usepackage{alltt}

\newcommand{\bfa}{{\mathbold a}}
\newcommand{\bfb}{{\mathbold b}}

\newcommand{\bfe}{{\mathbold e}}
\newcommand{\bff}{{\mathbold f}}

\newcommand{\bfl}{{\mathbold l}}
\newcommand{\bfm}{{\mathbold m}}
\newcommand{\bfn}{{\mathbold n}}

\newcommand{\bft}{{\mathbold t}}
\newcommand{\bfu}{{\mathbold u}}

\newcommand{\bfx}{{\mathbold x}}

\newcommand{\bfz}{{\mathbold z}}

\newcommand{\bfA}{{\mathbold A}}

\newcommand{\bfC}{{\mathbold C}}

\newcommand{\bfI}{{\mathbold I}}

\newcommand{\bfL}{{\mathbold L}}

\newcommand{\bfN}{{\mathbold N}}

\newcommand{\bfV}{{\mathbold V}}

\newcommand{\bfX}{{\mathbold X}}

\newcommand{\beq}{\begin{equation}}
\newcommand{\eeq}{\end{equation}}
\newcommand{\beqs}{\begin{eqnarray}}
\newcommand{\eeqs}{\end{eqnarray}}
\newcommand{\beql}{\begin{equation} \label}

\newcommand{\bfchi}{\mathbold{\chi}}

\newcommand{\bfalpha}{\mathbold{\alpha}}